\documentclass[aps,pre,reprint,floatfix]{revtex4-2}

\usepackage{graphicx}
\graphicspath{{Images_paper/}}
\DeclareGraphicsExtensions{.pdf,.png}

\usepackage{grffile}

\usepackage{xcolor}
\usepackage{amsmath}
\usepackage{amsthm, amssymb, amsfonts, amsbsy}
\usepackage{mathtools} 
\usepackage{bm}
\usepackage{mathrsfs}  
\usepackage{tipa} 

\usepackage{tikz}
\usepackage{tikz-3dplot}
\usepackage{ragged2e}

\usepackage[T1]{fontenc}
\usepackage[utf8]{inputenc}

\usepackage{textcomp}

\usepackage[english]{babel}

\newcommand{\beq}{\begin{equation}}
\newcommand{\eeq}{\end{equation}}
\newcommand{\beqa}{\begin{eqnarray}}
\newcommand{\eeqa}{\end{eqnarray}}
\newcommand{\bem}{\begin{math}}
\newcommand{\eem}{\end{math}}

\usepackage{bm}

\newcommand{\bmG}{{\bm G}}
\newcommand{\bmk}{{\bm k}}

\newcommand{\bmJ}{{\bm J}}

\newcommand{\bmV}{{\bm V}}

\newcommand{\bmq}{{\bm q}}
\newcommand{\bmQ}{{\bm Q}}
\newcommand{\bmD}{{\bm D}}

\newcommand{\bfr}{{\bf r}}

\newcommand{\bfk}{{\bf k}}

\newcommand{\bff}{{\bf f}}

\newcommand{\bfF}{{\bf F}}
\newcommand{\bfV}{{\bf V}}

\DeclareMathOperator{\Tr}{Tr}

\newcommand{\aver}[1]{\left\langle {#1}\right\rangle}

\newcommand\bnabla{\boldsymbol{\nabla}}

\newcommand{\abs}[1]{\lvert #1 \rvert}

\usepackage[capitalise]{cleveref}
\crefformat{equation}{Eq. (#2#1#3)} 
\Crefformat{equation}{Equation (#2#1#3)} 
\crefrangeformat{equation}{Eqs. (#3#1#4--#5\crefstripprefix{#1}{#2}#6)} 
\Crefrangeformat{equation}{Equations (#3#1#4--#5\crefstripprefix{#1}{#2}#6)} 
\crefmultiformat{equation}{Eqs. (#2#1#3)}{ and~(#2#1#3)}{, (#2#1#3)}{ and~(#2#1#3)} 
\Crefmultiformat{equation}{Equations (#2#1#3)}{ and~(#2#1#3)}{, (#2#1#3)}{ and~(#2#1#3)} 

\usepackage{xcolor}

\begin{document}


\title{Quantifying cell shape and density fluctuations in  epithelial tissue \textit{in vivo}} 

\author{J. Turley$^{1}$}
\author{H. Weavers$^2$}
\author{P. Martin$^2$}
\author{I.V. Chenchiah$^{3}$}
\author{T. B. Liverpool$^{3}$}

\affiliation{$^1$Mechanobiology Institute, National University of Singapore - Singapore, 117411}

\affiliation{$^2$School of Biochemistry, University of Bristol - Bristol BS8 1TW, UK}

\affiliation{$^3$School of Mathematics, University of Bristol - Bristol BS8 1UG, UK}

\date{\today}

\begin{abstract} 
Controlling changes in cell shape are crucial for many biological processes, such as tissue development and wound healing. Tissues typically are heterogeneous with a variety of cell shapes and sizes. Of particular interest are local deviations from an average cell shape and size. These fluctuations may extend and transmit across tissues, potentially offering valuable insights into tissue characteristics such as variations in effective "stiffness" or rigidity. In this study, we present a theoretical framework that captures the dynamics of epithelial cell shapes within tissue, incorporating both their average behaviour and fluctuation patterns. We model cells as interacting soft ellipsoids of varying size and aspect ratio. Coarse-graining our model, we obtain a set of continuum stochastic differential equations from which we derive spatial-temporal correlation functions. These correlation functions fit with our experimental data from the developmental process of the \textit{Drosophila} pupal wing. From the correlation functions, critical parameters representing active cell shape changes and effective tissue "stiffness" can be determined.
\end{abstract}

\maketitle
\section{Introduction}

During development, epithelial tissues undergo morphological changes that alter their packing, shape, and mechanical properties \cite{Tetley2019TissueHealing, Lenne2022SculptingTransitions, Mongera2018AElongation, rustarazo-calvo_adhesion-driven_2026, staddon_cell_2026, fuhrmann_active_2024}. Hence, these tissues have attracted chemists, physicists, and mathematicians to model these complex systems \cite{Tetley2018TheFluidity, Hirashima2017CellularMorphogenesis, leong_critical_2025, turley_iscience_2022}. Two broad approaches are to simulate each individual cell or to study a  coarse-grained model of the system using continuum equations. On the cellular scale, the vertex and cellular Potts models have both been used to study the dynamics of solid/fluid-like behaviours, the development of \textit{the Drosophila} wing imaginal disc, and wound healing \cite{Tetley2018TheFluidity, tozluolu_planar_2019, manning_rigidity_2024, park_unjamming_2015, leong_critical_2025, gomez_highly_2024}. In contrast, continuum models reduce complexity and hence are able to model key large scale, long-time characteristics of the tissue. They have been used to model tissue folding, buckling and the development of the \textit{Drosophila} pupal wings \cite{Etournay2015InterplayWing, ishihara_cells_2017, murisic_discrete_2015, andralojc_dynamics_2026, ioratim-uba_exotic_2025}. Cellular scale information (in the form of deviations from the local average) is typically ignored in tissue continuum models. Here we aim to augment the power of continuum models with cell scale information by incorporating local fluctuations. We will do this by coarse-graining  a microscopic model in which cells are parametrised as interacting ellipsoids of varying size and aspect ratio. We show how these local deviations from  tissue scale properties can give us important information about the processes governing their behaviour.

\begin{figure}
\centering

	\includegraphics[scale=1]{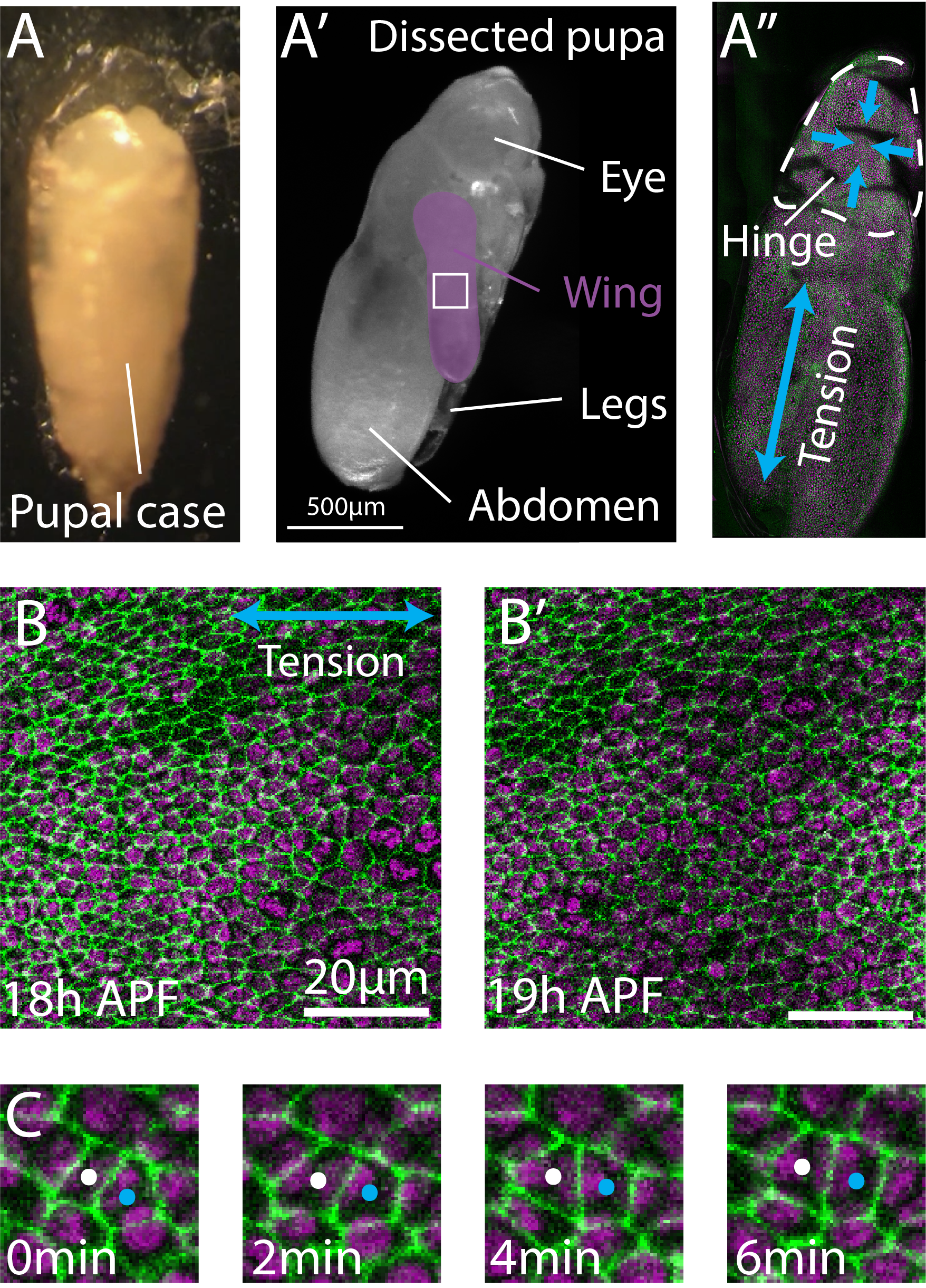} 
	\caption{\justifying \small \sl \textbf{\textit{Drosophila} pupal wing tissue and cell shape fluctuations.} A) a pupa inside its puparium. A') Bright field image of an 18hour APF (after puparium formation) pupa with the wing highlighted in magenta. B-B') Snap shots from a movie of the epithelium at timepoints indicated. \textit{E-cadherin-GFP} (green) shows cell boundaries and \textit{Histone2-RFP} (red) labels cell nuclei. C) Epithelial cells with fluctuating boundaries over time. } \label{fig:diagram}
\end{figure}

In this paper, we model the epithelium of the developing \textit{Drosophila} pupal wing (Fig. \ref{fig:diagram} A, B), an almost flat 2D surface. We investigate the tissue from 18 hours after puparium formation (APF) to 21 hours APF \cite{turley_deep_2024-1, Weavers2016SystemsGradient, turley_deep_2024}. During this period, the hinge region is condensed, stretching the wing blade where we will image \cite{Etournay2015InterplayWing, paci_forced_2021, athilingam_mechanics_2021}. This is a complex system with structures such as veins forming and is highly dynamic \cite{Etournay2015InterplayWing}.
In wing tissue, it is well documented in the literature that cells elongate along the orientation of the global tension, which is generated by the contraction of the hinge \cite{Etournay2015InterplayWing, paci_forced_2021, athilingam_mechanics_2021}. We will focus on studying how fluctuations in cell shape and size (Fig. \ref{fig:diagram} C) from this mean cell stretching evolve over time and how we use these to gain information about the mechanical properties of the tissue. We will model the cells as interacting soft ellipsoids able to change their area, elongation, and orientation. Starting from microscopic equations of motion, we coarse-grain the system to obtain stochastic continuum equations governing fluctuations in cell shape and density.

Our reductionist approach allows us to model the fluctuations in this system using only the leading-order terms and relatively few parameters. Even with these simplifications, we are able obtain a reasonably good fit of the experimental data to the model. These parameters may provide insight into how tissue-scale mechanical properties are linked to microscopic cell–cell interactions.

Finally, we test the framework on perturbed tissues with altered mechanical properties. We analyse laser-wounded tissues after wound closure, where local changes in mechanical properties such as fluidity have been reported \cite{tetley_tissue_2019}. We also consider genetic perturbation of JNK signalling, which is known to reduce cell shape remodelling during wound healing and delay closure \cite{turley_deep_2024, lee_cdc37_2019, Tafesh-Edwards2020JNKHomeostasis, weavers_injury_2019}. We hypothesise that these perturbations will lead to measurable changes in the model parameters inferred from fluctuations in cell shape and density.

\section{Experiment, imaging and analysis}

The \textit{Drosophila} pupal wing has a flat 2D surface with a single layer of epithelial cells \cite{Etournay2015InterplayWing, athilingam_mechanics_2021}. In addition to high-definition imaging and a variety of genetic tools, it is an ideal system to study tissue dynamics during morphogenesis and wound healing \cite{Weavers2018Long-termPupae, Weavers2016SystemsGradient, turley_deep_2024-1, Etournay2015InterplayWing}. To image the wing, we dissect the pupae removing the opaque outer casing (puparium) and place the entire pupae in an imaging dish with the wing in contact with the glass \cite{Weavers2018Long-termPupae}. They can now be imaged on the confocal microscope with pupae expressing \textit{E-cadherin-GFP} (to label cell boundaries) and \textit{Histone2-RFP} (for nuclei) \cite{turley_deep_2024-1}.

In previous studies, we have developed an image analysis pipeline to extract quantitative data from raw confocal images \cite{turley_deep_2024-1}. Here we will be using a deep learning algorithm to detect cell boundaries from these images via the \textit{E-cadherin-GFP} channel. After the cell boundaries have been located, we fit them with a polygon that can be used to measure the properties of the cell shape \cite{Olenik2023FluctuationsTissue}. We can use the cell boundaries to quantify cell shape and density. We label the nuclei to track the average movement in the tissue, allowing us to account for the tissue drift when obtaining correlation functions \cite{Tinevez2017TrackMate:Tracking}.

In this study, we began by investigating the tissue properties of fly wings throughout the pupal development period from 18 to 21 hours APF in the wing tissue. In the experiments, an image is taken every 2 minutes over the 3 hours. We are able to fit the parameters of our model to the experimentally measured correlation functions \cite{Etournay2015InterplayWing, athilingam_mechanics_2021}. 

\subsection{Cells as soft ellipsoids}

\begin{figure}[!ht]
\centering

	\includegraphics[scale=1]{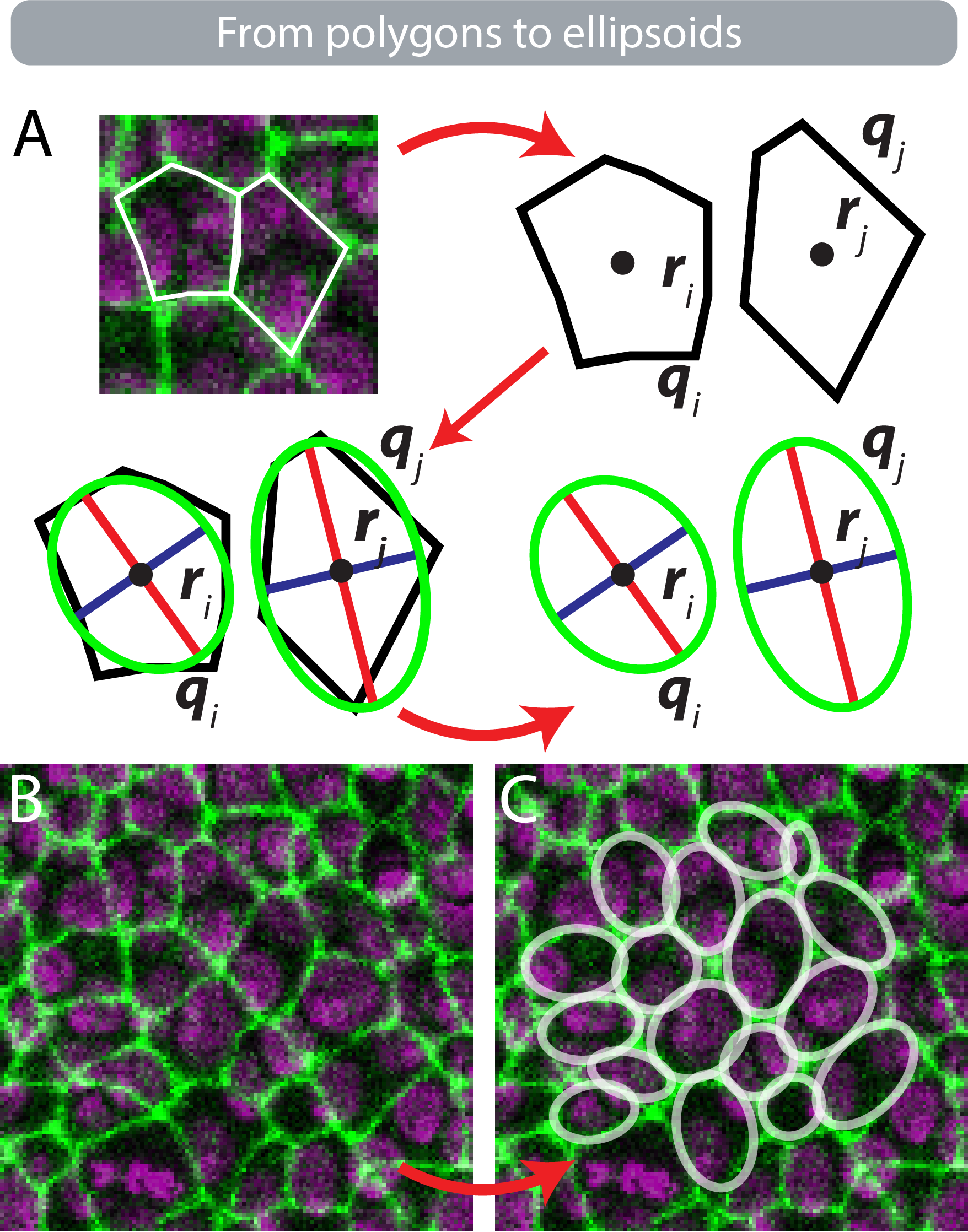} 
	\caption{\justifying \small \sl \textbf{Epithelial cells as soft ellipsoids.} A) Cell shapes can be approximated by polygons which we  subsequently map to ellipses. Each ellipse can be assigned a position (geometric centre) as well as a long (red) and short (blue) axis computed via a $q$-tensor evaluated from the polygonal cell shape. B) Zoom in of wing tissue. E-cadherin-GFP (green) shows cell boundaries and Histone2-RFP (red) labels cell nuclei. A) Epithelial cells with white overlay of ellipsoids.} \label{fig:diagram q-tensor}
\end{figure}

In this system, cells are elongating along the P/D axis (Proximal-Distal axis) of the wing with fluctuations in their shapes around this mean behaviour. Therefore, we will model the cells as soft ellipsoids that can change their elongation and orientation (Fig. \ref{fig:diagram q-tensor} A). Cells also become smaller, slowly increasing in density over the 3 hours we image the tissue. Therefore, we also allow them to change their area as well. We can give each cell in the tissue a position and shape. The position, $\bfr_i= (x_i, y_i)$ is the location of its geometric centre. The shape is encoded by a 2nd rank tensor, which we call the $q$-tensor, $\bmq_i$ that measures the deviation of the cell-shape from circular. For a circular cell, the $q$-tensor is identically zero and it is non-zero for an elongated cell which can be visualised as an ellipse. The eigenvectors of the $q$-tensor give the major (corresponding to the largest eigenvalue) and minor axis (corresponding to the smallest eigenvalue). The position of a cell is defined as the centroid of the cell, and the q tensor is a 2nd moment of the integral over the area of the cell that contains information about the elongation and orientation of an individual cell (Fig. \ref{fig:diagram q-tensor} B). The area and the $q$-tensor for each cell are defined as: 

\begin{align}
    A_i &= \iint_A   \,dx\,dy, \label{eq:c} \\
    \bmq_i &= \frac{1}{A_i^2}\iint_A \begin{pmatrix} \frac{1}{2}(x^2-y^2) & xy \\ xy & \frac{1}{2}(y^2-x^2) \end{pmatrix}\,dx\,dy, \nonumber \\
    \bmq_i &= q_i^{(0)}\left( \begin{array}{cc} \cos{2\theta_i} & \sin{2\theta_i} \\ \sin{2\theta_i} & - \cos{2\theta_i} \end{array} \right) = \left(\begin{array}{cc} q_i^{(1)} & q_i^{(2)} \\ q_i^{(2)} & - q_i^{(1)} \end{array} \right). \quad \nonumber
\end{align}
where $x,y$ in the integrands are measured w.r.t. to the centre of cell $i$, $A_i$ is the area of the cell $i$ and $\bmq_i$ is its $q$-tensor. The $q$-tensor is a second rank symmetrical matrix which can be decomposed into $q^{(0)}_i$, the cell's elongation (difference between major and minor axis length) and $\theta_i$, its orientation \cite{Olenik2023FluctuationsTissue}. These are calculated for every cell at every time point. More than 1,000,000 cell-level measurements were collected in this study.

\begin{figure}[!ht] 
\centering

	\includegraphics[scale=1]{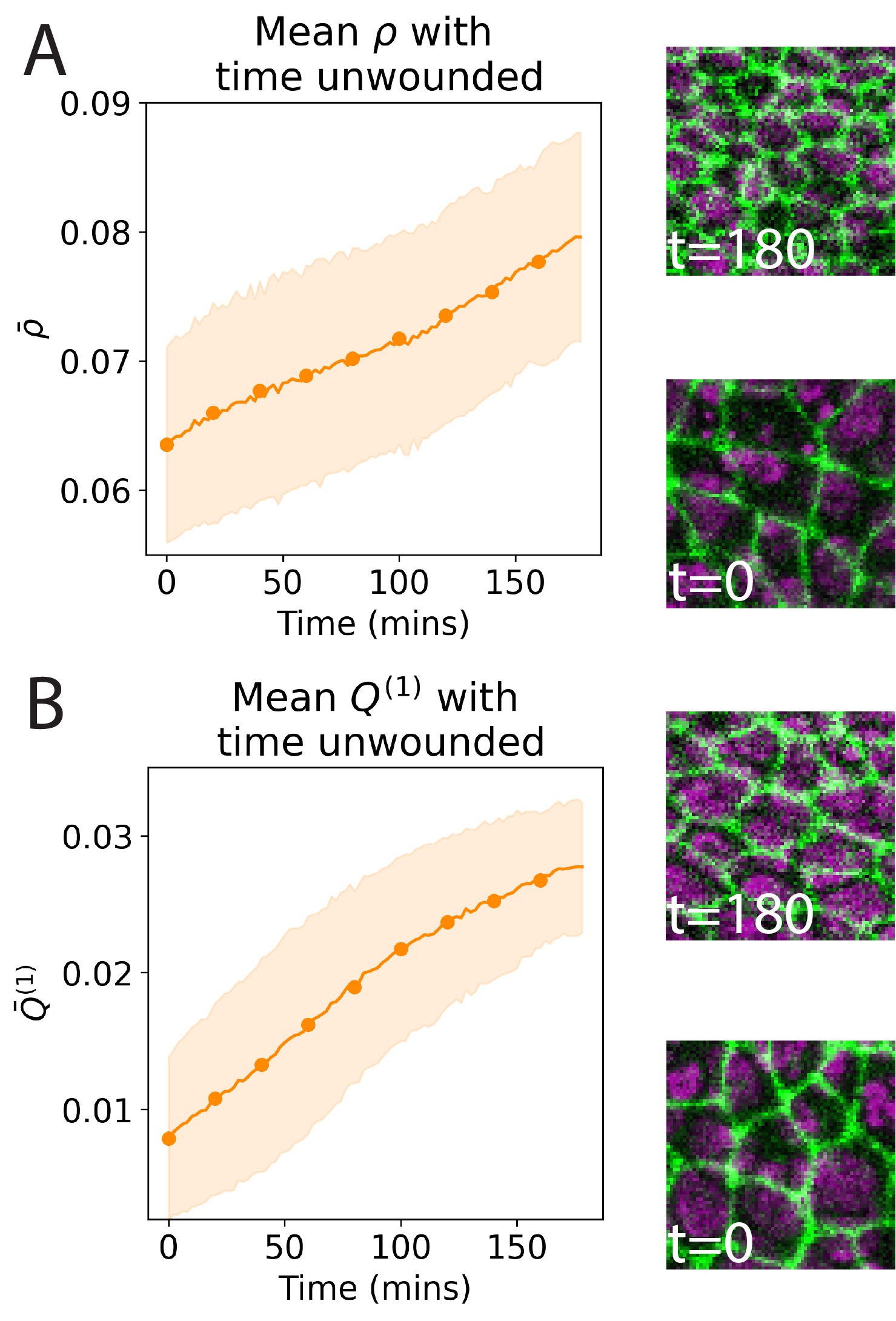} 
	\caption{\small \justifying \sl \textbf{Mean of cell density and elongation over time.} A) The mean $\rho$ over time with orange shaded region, showing the standard deviation. B) The mean $Q^{(1)}$ over time with orange shaded region, showing the standard deviation.} \label{fig:mean behaviour}
\end{figure}

Now we can compute the mean cell properties over time by defining the density $\rho = 1/\langle A \rangle$ and the mean $q$-tensor $\bar{\bmQ} = \langle \bmq \rangle$. In this stage of development, cells are elongated along the proximal-distal (P/D) axis of the wing. We will define the P/D axis as the orientation of the average cell. This axis has the same orientation as the global tissue tension \cite{Etournay2015InterplayWing, athilingam_mechanics_2021}. This leads us to simplify the components of $\bar{\bmQ}$ with $\bar{Q}^{(1)} = \bar{Q}^{(0)}\cos(\theta) = \bar{Q}^{(0)}$ and $\bar{Q}^{(2)} = \bar{Q}^{(0)}\sin(\theta) = 0$ by definition \cite{Olenik2023FluctuationsTissue}. Now when $\bar{Q}^{(1)}>0$ cells are on average elongated along the P/D axis and when $\bar{Q}^{(1)}<0$ they are perpendicular to it, along the anterior-posterior axis (A/P axis). When $\bar{Q}^{(2)}>0$ cells are on average elongated along the axis at 45$^o$ to the P/D axis and when $\bar{Q}^{(2)}<0$ elongated along the axis at 135$^o$ to the P/D axis. We can thus determine the mean cell properties for each time point. We found that during the period of observation that both $\rho$ and $\bar{Q}^{(1)}$ increasing linearly with time, thus the cells are increasingly aligned along the P/D axis (Fig. \ref{fig:mean behaviour}A, B). The increase in density is mainly due to cell division that occurs in tissue during this stage of development \cite{Etournay2015InterplayWing, turley_deep_2024-1}.

Although this information is helpful for understanding the tissue, it gives only a general overview. To gain a richer understanding of the system, we look at the individual deviations of cells from this average. We define:
\begin{align*}
    \delta \bmq_i &= \bmq_i - \Bar{\bmQ}, \\
    \delta \bmq_i &= \left( \begin{array}{cc} \delta q^{(1)} & \delta q^{(2)} \\ \delta q^{(2)} & - \delta q^{(1)} \end{array} \right),
\end{align*}
where $\delta \bmq_i$ are the deviations from the mean $q$-tensor. The same can be done with the local density $\delta \rho (\bfr) = \rho - 1/\langle A \rangle$ where we locally average the area of cells in a box $10\times10\mu m$ centred on $\bfr$. 

\section{Theory}

We start with microscopic equations of motion for the positions of the cell centres, $\bfr_i(t) $,  and cell shape ($q$-tensor), $\bmq_i(t)$. We then use them to generate equations for fluctuating fields measuring local cell density and cell shape  by averaging and coarse-graining. 

We will describe the large scale behaviour by the coarse-grained fields for cell density: $\hat{\rho}(\bfr,t)$, and cell shape: $\hat{\bmQ} (\bfr,t)$ defined as
\begin{align*}
\hat\rho(\bfr, t) &= \sum_i \delta (\bfr_i(t)- \bfr) =\sum_i \hat \rho_i (\bfr,t) \, ,  \\
\hat\rho(\bfr, t) \hat{\bmQ} (\bfr, t) &= \sum_i \bmq_i (t)\delta (\bfr_i(t)- \bfr) =\sum_i \bmq_i \hat \rho_i \, ,
\end{align*}
where $\hat \rho_i (\bfr,t)=  \delta (\bfr_i (t) - \bfr)$.
In a previous study, we have also calculated the dynamics of cell shape polarisation~\cite{Olenik2023FluctuationsTissue}, i.e. breaking the nematic symmetry. In our pupal wing system, we found that this was negligible and on average zero hence we have not included it here.

\subsection{Microscopic equations}

Since one does not have all the biochemical information required to completely determine the form of the microscopic equations for the cell positions (their geometric centres) and ellipsoidal shapes ($q$-tensor) in the tissue being studied, we will proceed with a {\em specific} simplified model with self- and pairwise interactions. We will then try to infer the form of the interactions and how they influence our tissue dynamics using experimental data both on their typical large scale properties {\em and} the fluctuations around them. The framework for generic microscopic models is outlined in an appendix.

We define the microscopic equations for the shape and positions with deterministic contributions leading to the average properties, pairwise interactions, and fluctuations (whose origin in general will be both mechanical and chemical)
\begin{align} 
\zeta_q \partial_t q_{i\alpha\beta} &= A (q_{i\alpha\beta} - \bar{Q}_{\alpha\beta}) - \frac{\partial U}{\partial q_{i\alpha\beta}} + \xi_{i{\alpha\beta}}^{q},  \\
\zeta_r \partial_t r_{i\alpha} &= \zeta_r \bar{V}_{\alpha} - \frac{\partial U}{\partial r_{i\alpha}} + \xi_{i \alpha}^{r}.
\end{align}
Here $q_{i,11} = - q_{i,22} = q_{i}^{(1)}$ and $q_{i,12} = q_{i,21} = q_{i}^{(2)}$. 
The term proportional to $A$ is the active tendency of alignment of the cell shapes with the mean, $\bar \bmQ (t)$. $\zeta_q$ and $\zeta_r$ are local "frictions" for the shape and the positions. We define a tissue velocity, $\bar{V}_{\alpha}$, the mean cell centre velocity averaged over all cells.  The pairwise interactions are taken as due to an interaction potential $U (\{ \bfr_i\},\{ \bmq_i\} ) $: 
\begin{equation*} 
U = \sum_{i\neq j} W(\bfr_i - \bfr_j) \Tr(\bmq_i \bmq_j).
\end{equation*}
This is a significant constraint which can be broken by on-equilibrium active elements like cells. However extending it to more general scenarios (like non-reciprocal interactions)  are relatively straight-forward with this framework in place. $W(\bfr_i - \bfr_j)$ is an "energy" of the shape interactions between cells. The interaction depends on both their shapes and alignment $\Tr(\bmq_i \bmq_j)$ and is an even function. The interaction energy is minimised when cells are aligned. We do not prescribe the form of $W$ at this point: we will calculate its properties and how it influences the tissue dynamics using experimental data. We set $\zeta_q=1$ and $\zeta_r=1$. All white noises $(\xi^m_{in...})$ have zero mean and are uncorrelated with each other. $\bm{\xi}^q_i$ and $\bm{\xi}^r_i$ are the uncorrelated white noises of cell $i$ for changes in shape and position respectively, 
where 
\begin{align}
\langle \xi_{i {\alpha}}^{r}(t) \xi_{j {\alpha}'}^{r}(t') \rangle &= 2 D_{r} \delta_{\alpha \alpha'} \delta_{ij} \delta(t' - t) \; , \notag \\ 
\langle \xi_{i \alpha\beta}^{q}(t) \xi_{j\alpha'\beta'}^{q}(t') \rangle &= 2 D^q_{\alpha \beta} {\delta_{\alpha\alpha'}}\delta_{\beta\beta'} \delta_{ij} \delta(t' - t) \; , \label{eq:noise_q}
\end{align}
with
\beq
{\bm D}^q = \left( \begin{array}{cc} D_{1} & D_{2} \\ D_{2} & -D_{1}   \end{array}\right).
\label{eq:noise_qmat}\eeq
$D_r$ measures the diffusion of the cell centres and ${\bm D}_q$ measures the ``diffusion" of cell shapes.
We make the simplifying assumption that in the dense tissue, the interaction forces balance and  $\bnabla_i U =\sum_j \bnabla W(\bfr_{ij})\Tr (\bmq_i \bmq_j) \simeq 0$. A further simplification is suggested by the experimental observation that motion of the cells in the tissue relative to their neighbours is small so that $\bar{V}_{\alpha}$ can be taken constant \cite{turley_deep_2023}.

\subsection{Coarse-graining the microscopic equations}

We will coarse-grain these equation to generate a continuum model in terms of the local shape field, $\hat\bmQ (\bfr,t)$ and local density field, $\hat\rho(\bfr,t)$.

\subsubsection{Continuum equations for shape}

From the definition of the fields, it follows that 
\beq
\partial_t (\hat \rho \hat \bmQ) = \sum_i \partial_t \bmq_i (t)\delta (\bfr_i(t)- \bfr) - \sum_i \bmq_i (t) \partial_t \bfr_i \cdot \bnabla \delta (\bfr_i(t)- \bfr) \; . \nonumber
\eeq
Applying the It\^o-calculus, see Appendix, we get
\beqa
\mathrm{D}_t (\hat \rho \hat \bmQ_{})  &=&   D_r \bnabla^2  \left(\hat \rho \hat \bmQ\right) + A \left[(\hat \rho  \hat \bmQ)  - \hat \rho \bar{\bmQ}_{} \right] \nonumber \\ &&  - \sum_i \left.  \frac{\partial U}{\partial \bmq_{i}}  \right. \delta (\bfr_i(t)- \bfr) + \bm{\xi}^Q (\bfr,t) \; , \label{eq:dotQExpand} \eeqa
where
\begin{align*}
\mathrm{D}_t (\hat \rho \hat \bmQ_{}) &=\partial_t (\hat \rho \hat \bmQ) + \bar \bfV \cdot \bnabla  (\hat \rho  \hat \bmQ), \\ 
\bm{\xi}^Q (\mathbf{r},t) &= - \sum_i \bnabla \left(\hat \rho_i \, {{\bmq}_i}\right)
\cdot \bm{\xi}^r_i 
+  \sum_i \left . \hat \rho_i  \,  \right.
\bm{\xi}^q_i \, .
\end{align*}

\subsubsection{Continuum equation for density}

We can keep track of cell sizes by mapping to the local cell density, $\hat \rho (\bfr,t)$. From the definitions of the fields and application of the It\^o calculus  we get an equation for the density: 
\beqa
\mathrm{D}_t \hat \rho  &=&   D_r \bnabla^2 \hat \rho    - \sum_i \left.   \bnabla_i U  \right. \delta (\bfr_i(t)- \bfr) + \hat {\bm \xi} (\bfr,t) \; , \nonumber \\  && \eeqa
where
\begin{align*}
\mathrm{D}_t \hat \rho  &=\partial_t \hat \rho + \bar \bfV \cdot \bnabla  \hat \rho, \\
\hat {\bm \xi} (\mathbf{r},t) &= - \sum_i \bnabla \hat \rho_i 
\cdot \bm{\xi}^r_i  \,   \, .
\end{align*}

With our simplifying assumptions above, this can be written as a continuity equation for the density of cells~\cite{dean1996langevin}:
\begin{equation*} 
\partial_t \hat \rho(\bfr, t) + \bm{\nabla} \cdot \hat \bmJ = 0,
\end{equation*}
where the flux $\hat \bmJ$ is given by:
\beq
\hat \bmJ =  \bar \bmV  \hat \rho - D_r \bnabla \hat \rho + \hat{\bm \xi} (\bfr,t),
\eeq
with
\begin{align*}
\langle \hat{\bm \xi} \rangle &= 0, \\
\langle \hat\xi_\gamma^{}(\bfr,t) \hat\xi_\epsilon^{}(\bfr',t') \rangle &= 2 D_r \hat \rho \; \delta_{\gamma\epsilon} \; \delta(t - t') \delta(\bfr- \bfr').
\end{align*}
The dynamics of the cell density is also  affected by (active) cell size changes. If the cell number is conserved, this can also be described by an additional local flux $\hat \bmJ_a$ given by
\beq
\hat \bmJ_a =  - D_a \bnabla \hat \rho + {\bm \xi}^a (\bfr,t), 
\eeq
with 
\begin{align*}
\langle {\bm \xi}^a \rangle &= 0, \\
\langle \xi_\gamma^{a}(\bfr,t) \xi_\epsilon^{a}(\bfr',t') \rangle &= m_a \; \delta_{\gamma\epsilon} \; \delta(t - t') \delta(\bfr- \bfr')\hat \rho,
\end{align*}
where $D_a,m_a$ are parameters to be fixed by data.

Next, we include the effect of cell division and cell death (apoptosis).
The divisions in the tissue are a source of cells, $d (\bfr,t)$  and apoptosis $a (\bfr,t)$ acts as a sink leading to a modified equation for the cell density. 
\begin{align} 
\partial_t \hat \rho(\bfr, t) + \bm{\nabla} \cdot \bmJ_{T}  = d - a \, ,
\end{align}
where the total flux, $\bmJ_{T}=(\hat \bmJ +\hat \bmJ_a)$. In our reductionist approach, we assume that these processes occur uniformly throughout the tissue, i.e. $a$ and $d$ are constant.

Finally, we note that the pupal wing is a developing tissue with cells destined to form structures such as veins. These can be modelled as  quenched (structural) density deviations, $\rho_s (\bfr)$ from the uniform mean density of cells in the wing. The quenched density  does not depend on time. The diffusive dynamics above then acts on the deviations from these quenched structural features, therefore we can write the total flux as follows:
\begin{align}
\hat\bmJ_T = \bar \bmV  \hat \rho - D \bm{\nabla} \hat \rho  + \bm{\xi}^{\rho} + \bm{\xi}_s (\bfr)\, , 
\end{align}
%
where $D=D_r+D_a$ and $\bm{\xi}^{\rho} =\bm{\hat\xi} + \bm{\xi}^{a} $.
The structural density deviations are driven by quenched fluctuations, $\bm{\xi}_s (\bfr)$, where
\beq
\bar \bmV \rho_s - D \nabla \rho_s + \bm{\xi}_s =0, \eeq
giving us the equation for cell density:
\begin{align} 
\mbox{D}_t \left( \hat\rho - \rho_s \right) = D\nabla^2 (\hat \rho - \rho_s(\bfr)) - \bm{\nabla} \cdot \bm{\xi}^\rho + d - a, \label{eq:diffuse density noise}
\end{align}
where $\mbox{D}_t f (\bfr,t) =  \partial_t f + \bar \bmV \cdot \bnabla f$.

We model the structural deviations from uniformity as correlated Gaussian fluctuations:
\begin{align}
\langle \rho_s (\bfr)  \rangle &= 0, \\
\left\langle  \rho_s (\bfr)  \rho_s (\bfr') \right\rangle &= F_s (\bfr - \bfr'), \label{eq:diffuse delta density structure}
\end{align}
to be parametrised by the experimental data.

\subsubsection{Local averaging and linearization}
We locally average cell properties to determine  smooth vector fields, $\rho(\bfr,t)$, $\bmQ(\bfr,t)$:
\beqa
\rho (\bfr,t) &=& \aver{\hat\rho(\bfr,t)} = \int d^2 r' \, K(\bfr-\bfr') \hat\rho(\bfr',t) \quad ,  \\  \rho (\bfr,t)\bmQ(\bfr,t) &=& \aver{\hat\rho(\bfr,t)\hat\bmQ(\bfr,t)} \nonumber \\ &=& \int d^2 r' \, K(\bfr-\bfr') \hat\rho(\bfr',t) \hat \bmQ(\bfr',t)
\quad , 
\eeqa
where $K(\bfr)$ is a finite-ranged kernel.

Next we linearise the equations by expanding about the uniform average density, ($\rho = \Bar{\rho} + \delta\rho$)  and mean shape, ($\bmQ = \Bar{\bmQ} + \delta\bmQ$)  , keeping terms up to linear order in the deviations. We also perform a long-wavelength expansion, keeping the leading order terms in  gradients for both shape, $\delta \bmQ (\bfr,t)$  and density, $\delta\rho (\bfr,t)$ deviations. Finally, we work in the frame of reference where $\bar\bmV=0$.

Substituting $\rho(\bfr, t)= \bar{\rho}  + (d - a)t + \rho_s(\bfr) + \delta\rho (\bfr, t)$, we get an equation for the (annealed) density deviations 
\begin{align} 
\partial_t \delta\rho(\bfr, t) &= D\nabla^2 \delta\rho(\bfr, t) - \bm{\nabla} \cdot \bm{\xi}^{\rho}(\bfr, t) \label{eq:diffuse delta density}
\end{align}
where
\beq
\langle {\bm \xi}^\rho \rangle =0\; , \quad  \langle \xi_\gamma^{\rho}(\bfr,t) \xi_\epsilon^{\rho}(\bfr',t') \rangle = m_\rho  \; \delta_{\gamma\epsilon} \; \delta(t - t') \delta(\bfr- \bfr') \; , 
\label{eq:rho-noise}\eeq 
with $m_\rho=2 D_r \bar \rho + m_a$.

The structural density deviations, $\rho_s$  given by  
\[
\left\langle \rho_s(\bfr) \right\rangle = 0,
\qquad
\left\langle \rho_s(\bfr)\rho_s(\bfr') \right\rangle
= F_s(\bfr-\bfr').
\]
are determined from the data. 

For the shape dynamics, we have $\bar Q^{(2)}=0$ and 
%
\begin{align} 
\partial_t \bar{Q}^{(1)}(t)&= 2b \bar\rho \bar{Q}^{(1)} \label{eq:bulkQ}\\
\partial_t \delta Q^{(1)}(\bfr, t) &=  [-A + 2 b \bar\rho + \left(D_r \delta_{\gamma\epsilon} +  \bar\rho L_{\gamma\epsilon} \right) \partial_{\gamma} \partial_{\epsilon}] \delta Q^{(1)}\nonumber\\
&+ \bar{Q}^{(1)} L_{\gamma\epsilon}\partial_{\gamma} \partial_{\epsilon} \delta \rho(\bfr, t) + \xi^{(1)}_{Q}(\bfr, t)\\
\partial_t \delta Q^{(2)}(\bfr, t) &= [-A + 2b \bar\rho + \left( D_r \delta_{\gamma \epsilon} +\bar\rho L_{\gamma\epsilon}\right) \partial_{\gamma} \partial_{\epsilon}]\delta Q^{(2)}  \nonumber \\
&+ \xi^{(2)}_{Q}(\bfr, t)
\end{align}
with $\langle \xi_{Q}^{(\alpha)}(\bfr, t) \rangle=0$ and 
\beq
\langle \xi_{Q}^{(\alpha)}(\bfr,t) \xi_{Q}^{(\beta)}(\bfr',t') \rangle = m_{Q}^{(\alpha)} \delta_{\alpha\beta} \delta(t' - t) \delta(\bfr' - \bfr) \quad , 
\label{eq:Q-alpha-noise}\eeq 
where $m_{Q}^{(\alpha)} = 2 D_{\alpha} \bar \rho $ with $\alpha,\beta  \in [1,2]$ and $D_\alpha$ as defined in eqn. (\ref{eq:noise_qmat}). Interactions between the cells in the bulk generates a passive isotropic stress (pressure) in the tissue, encoded by $b$. Similarly an  effective "stiffness" of the tissue, encoded by $L_{\gamma\epsilon}$,  is generated. $b,L_{\alpha\beta}$ are both parameters that must be fitted to the data. These parameters are obtained from integrals over the energy of shape interactions as shown below.

\begin{align} 
b &= - \int_{\mathbb{R}^2} d^2 r \, W(\bfr) \label{eq:b} \\
L_{\gamma\epsilon} &= -\int_{\mathbb{R}^2} d^2 r \, r_\gamma \, r_\epsilon W(\bfr) \label{eq:L}
\end{align}

We will assume that $L_{\gamma\epsilon} = L \delta_{\gamma\epsilon}$ and $b, L>0$. A simplification we have made is that we set the mean density to a constant $\bar \rho \approx \rho^* = 0.0713\mu m^{-2}$ this is the value averaged over time. Also, from the data and by definition $\bar{Q}^{(2)}=0$ for all $t$.  

Lastly, we used experimental data to measure the correlation between $\delta Q^{(1)}$ and $\delta \rho$. After normalisation we found no correlation (Fig. S1). This suggests that we can neglect the coupling to $\delta \rho$ in the equation for $\delta Q^{(1)}$: 
\begin{align} 
\partial_t \delta Q^{(1)}(\bfr, t) &= [-A + 2b \rho^* + \left (D_r \delta_{\gamma\epsilon} + \rho^*L_{\gamma\epsilon} \right) \partial_{\gamma} \partial_{\epsilon}]\delta Q^{(1)} \nonumber \\
&+ \xi^{(1)}_{Q}(\bfr, t)
\end{align}
while the $\delta Q^{(2)}$ equation remains unchanged, 
\begin{align}
\partial_t \delta Q^{(2)}(\bfr, t) &= [-A + 2b \rho^* + \left( D_r \delta_{\gamma \epsilon} +\rho^* L_{\gamma\epsilon}\right) \partial_{\gamma} \partial_{\epsilon}]\delta Q^{(2)}  \nonumber \\
&+ \xi^{(2)}_{Q}(\bfr, t)
\end{align}

\subsection{From correlations to large scale behaviour}

We introduce the Fourier transforms, 
\beqa
[\delta \tilde{Q}^{(\alpha)} (\bfk,t) , \tilde{\xi}_Q^{(\alpha)} (\bfk, t)] &= & \int d^2 r \, e^{i \bfk \cdot \bfr} \,  [\delta Q^{(\alpha)}(\bfr, t) , \xi^{(\alpha)}_{Q}(\bfr, t)] \, , \, \nonumber \\ 
 \delta \tilde{\rho} (\bfk,t) &= &  \int d^2 r \; e^{i \bfk \cdot \bfr} \,  \delta \rho (\bfr, t) \, , \, \nonumber \\ 
 \tilde{\xi}_\alpha^\rho (\bfk,t) & = &   \int d^2 r \, e^{i \bfk \cdot \bfr} \,  {\xi}^\rho_\alpha  (\bfr, t) \, , \, \nonumber 
\eeqa
where $\alpha \in [1,2]$.
%
We thus obtain the following equations: 
\begin{align} 
\partial_t \delta \tilde{Q}^{(\alpha)}(\bfk, t) &= [-B - (D_r - \rho^*L) k^2]\delta \tilde{Q}^{(\alpha)}(\bfk, t) + \tilde{\xi}_{Q}^{(\alpha)}(\bfk, t) \\
\partial_t \delta \tilde{\rho}_{}(\bfk, t) &= - D k^2 \delta\tilde{\rho}_{}(\bfk, t) + i k_\gamma \tilde{\xi}_{\gamma}^{\rho}(\bfk, t) 
\end{align}
with $B = A - 2b\rho^*$ and ($B>0$) and $k=|\bfk|$. These  1st order inhomogeneous ODEs have general solutions 
\begin{align} 
\delta \tilde{Q}^{(\alpha)} (\bfk, t) &= \int_{-\infty}^t d\tau \, \tilde{\xi}^{(\alpha)}_{Q}(\bfk, \tau) \, e^{[A - 2\rho^*b + (D_r +\rho^* L) k^2](\tau -t)} \label{eq:Cqq} \\
\delta\tilde{\rho}(\bfk, t) &= \int_{-\infty}^t d\tau \,  i k_\gamma \, \tilde{\xi}_{\gamma}^{\rho}(\bfk,\tau) e^{-D k^2(t-\tau)} \quad . \label{eq:crhorho}
\end{align}

Defining the set of {\em space-time} correlation functions for the dynamic variables,
\begin{align}
    C^{\alpha\beta}_{qq}(R, T) &= \langle \delta Q^{(\alpha)}(\bfr, t) \delta Q^{(\beta)}(\bfr', t') \rangle \quad , \\
    C^{}_{\rho\rho}(R, T) &= \langle \delta \rho (\bfr, t) \delta \rho (\bfr', t') \rangle \quad , 
    \end{align}
where $R=|\bfr-\bfr'|$ and $T=|t-t'|$, we obtain the following expressions from the general solutions above 
%
%
\begin{align}
  C^{}_{\rho\rho}(R, T) &= \frac{m_{\rho}}{32 D^2\pi^3 T} e^{- \frac{R^2}{4 D T}} \label{eq:corr_rho} \\
  C^{\alpha \beta}_{qq}(R, T) &= \delta_{\alpha \beta} \frac{m_{Q}^{(\alpha)}}{32\pi^3} \int_0^\infty dk \frac{k J_0(kR)}{B + L^* k^2} e^{-(B + L^* k^2)T} \label{eq:corr_q}
\end{align}
where $L^*=D_r + \rho^*L$.

Also of interest is the {\em spatial} structural density correlation function that is independent of time
\beq
    C_{ss} (R) = \langle \rho_s(\bfr) \rho_s(\bfr') \rangle
\eeq
which we determine empirically from our pupal wing data. Experimentally we measure a total density correlation function
\beq
C^{tot}_{\rho\rho}(R, T) = C_{ss} (R) + C^{}_{\rho\rho}(R, T) \quad .
\eeq

In the next sections, we will fit these to our experimental data to determine the parameters, $m_\rho, m_Q^{(\alpha)}, D, D_r,L,A,b$. In matching  the shape correlation functions, $C_{qq}^{\alpha\beta}$, between theory and experiment, we do not coarse-grain the shape field in the experimental data before averaging but evaluate the spatio-temporal correlations over all pairs of individual cells.

\subsection{Fluctuations, correlations and  experiments}

\begin{figure}[!ht] 
\centering
	\includegraphics[scale=1]{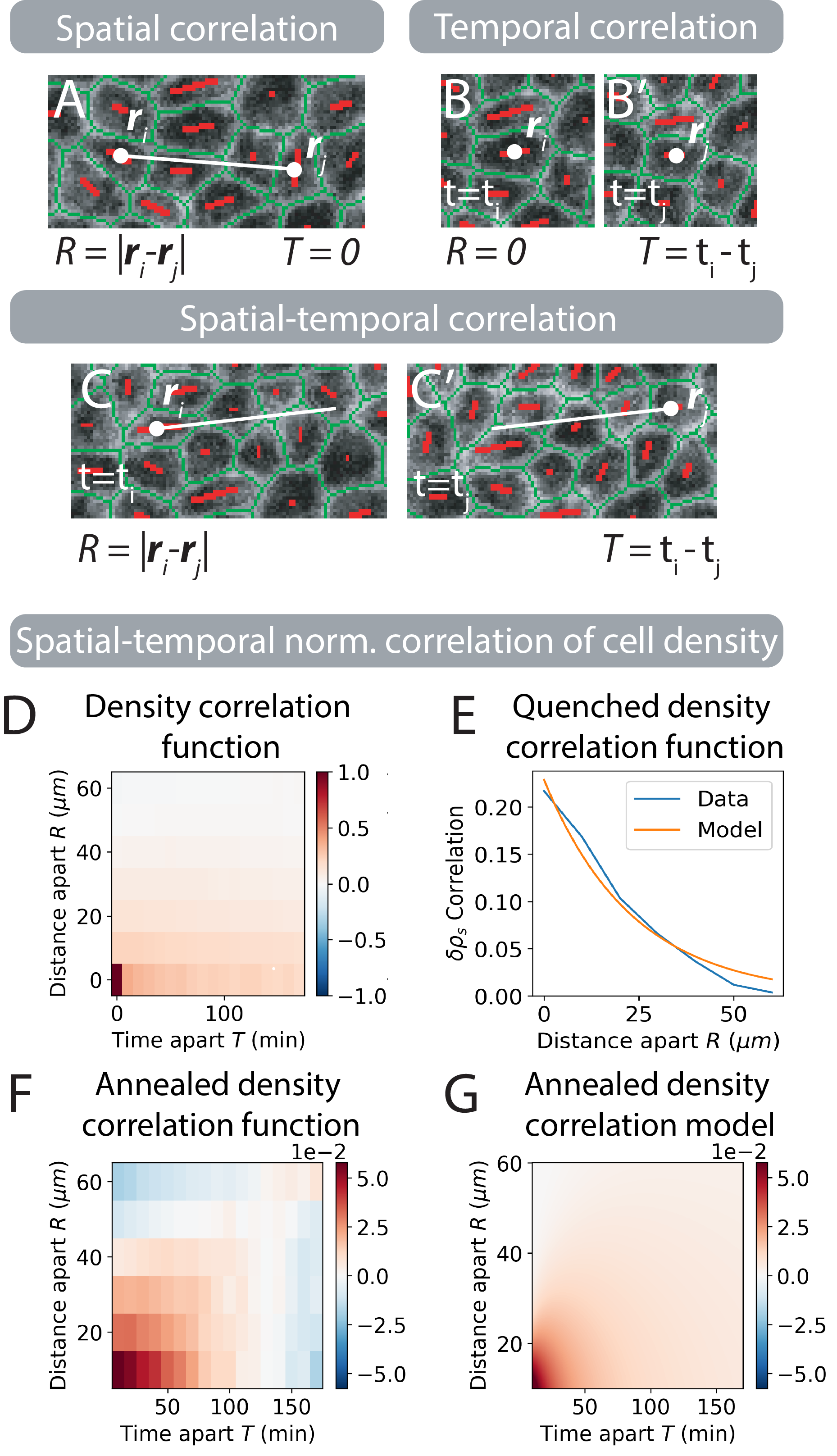} 
	\caption{\justifying \small \sl \textbf{Spatio-temporal density correlations functions from experiments matches theoretical models predictions.} A) Same time correlation, where $R$ > 0, separation between cells is non-zero, but evaluated at the same same time point. B) Same position correlation, where the time interval between events is non-zero, $T$ > 0 but $R$ = 0, position is fixed. C) Spatio-temporal correlation function diagram now varying both separation of positions $R$ > 0 and time-interval $T$ > 0. D) Heatmap of the space and time correlation of density fluctuations. Red regions indicate relatively high positive correlation between the two properties and blue areas negative correlation. White parts of the map show no correlation. E) Fitting model parameters to experimental data of the structural density deviations. F-G) Heatmaps of the annealed density correlations in space and time of experimental data and model with fitted parameters, respectively. } \label{fig:unw_corr}
\end{figure}

To compare the theory with the experimental data, we will need to calculate the spatial-temporal correlation functions of the fluctuations. Here we average all pairs of cells where $R < \abs{\bfr-\bfr'}<R+\delta R$ and $T < t' - t <T+\delta T$ are true. For the $q$-tensor correlations $\delta R = 2 \mu m$ and $\delta T = 2$ min. For local density deviations $\delta \rho$ larger bins are used ($\delta R = 10 \mu m$, $\delta T = 10$ min) due to the lower resolution of the local density. Fig. \ref{fig:unw_corr}A-C displays images of pairs of cells in space at the same time ($t=t'$ but $\bfr\ne \bfr'$), in time at the same position ($\bfr=\bfr'$ but $t \ne t'$) and in space-time, respectively. Evaluating correlation functions implies comparisons of the cells' $q$-tensors and sizes. The coarse-grained theory is only valid for length scales larger than cells; therefore, these correlation functions are evaluated for $R\geq10\mu m$. The typical cell is $\approx5\mu m$

\subsubsection{Fitting parameters of the model to pupal wing tissue}

We started by finding the parameters which can be determined from the averaged experimental data. From eqn. (\ref{eq:bulkQ}) it follows that
\begin{align} 
  \bar{Q}^{(1)}(t) &= \bar{Q}^{(1)}_0e^{2b\bar{\rho} t} \approx \bar{Q}^{(1)}_0 + 2\bar{Q}^{(1)}_0 b\bar{\rho} t \quad .
\end{align}
Taking a linear approximation of the solution and via a line of best fit to the experimental data we get a value for $b=0.0179\mu \mbox{m}^{2} \min^{-1}$. 

To study the mechanics of the tissue, we will compute a set of temporal-spatial correlation functions of the deviations in cell properties. These show us how fluctuations propagate throughout the tissue in space and time. If fluctuations decay slowly in space, this will indicate a stiffer, more rigid tissue, whereas rapid spatial  decay would show a more flexible, softer tissue. In addition, the magnitude of these shape/size fluctuations can be used as measures of  an ``{\em effective temperature}'' of the cell shapes/sizes. 

Using the correlation functions from the theoretical model, we can fit the parameters of the system. We begin with the structural density deviations which can be extracted as the {\em time-independent} part of $C_{\rho\rho}^{tot}$ (Fig. \ref{fig:unw_corr} D). We do not have a theoretical prediction for this: it is driven by upstream biochemical processes controlling development that we do not yet understand. 
Hence we fit to the simplest model of a correlation which decays with distance, an exponential decay over a correlation length, $\lambda$: 
\begin{align}
C_{ss}(R) = m_s \,  e^{-\frac{R}{\lambda}} \quad .
\end{align}
The best fit gives the magnitude of the quenched fluctuations as $m_s^{} = 4.3\times10^{-4}\mu \mbox{m}^{-4}$ and the decay length as $\lambda = 23 \mu$m. Fig. \ref{fig:unw_corr} E shows the data for $C_{ss} (R)$ and the best fit model.

Next, we extract the model parameters for the annealed density correlations, defined in eqn. (\ref{eq:crhorho}). They were determined using a 2D least squares method. This gave values of the diffusion constant and magnitude of the annealed fluctuations as $D = 7.4 \mu \mbox{m}^{2} \mbox{min}^{-1}$ and $m_{\rho } = 109 \mu \mbox{m}^{-4}$, respectively. The space-time correlations of the model and the data are plotted in Fig. \ref{fig:unw_corr} F, G, respectively.

\begin{figure}[!ht]
\centering

	\includegraphics[scale=1]{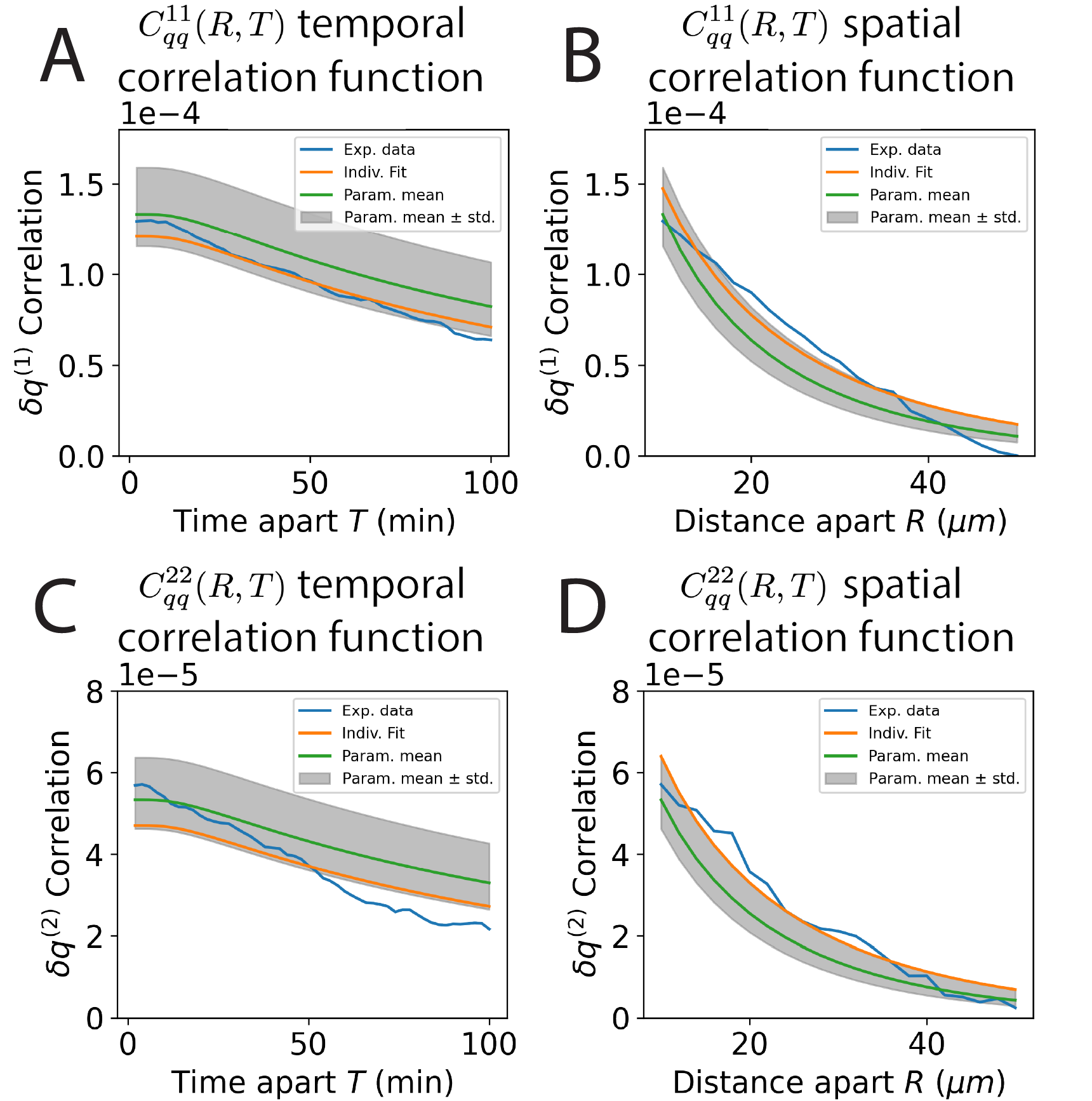} 
	\caption{\justifying \small \sl \textbf{Fitting model parameters to correlation of cell shape.} Correlations of cell shape for A, B) $\delta Q^{(1)}$ and C, D) $\delta Q^{(2)}$ for fixed $T$ and $R$ respectively. Graphs show the experimental data (blue), fitted parameters for individual cases (green), model using averaged parameters (orange) and the standard deviation of the averaged model parameters (gray).} \label{fig:dq_corr}
\end{figure} 

Finally we extract the parameters for the more complex $q$-tensor correlations. One complication arises from the numerical evaluation of  the integral in eqn. (\ref{eq:corr_q}). Using the least squares method on the full correlation functions in space and time becomes too computationally expensive. Therefore we made simplifications fitting these functions to the data for (fixed $T=0$, varying $R$) and (fixed $R=10$, varying $T$) for both $C^{11}_{qq}$ and $C^{22}_{qq}$ will give us values for the parameters that somewhat differ for each case, and we will average these to get the values that best fit all the data. This reductionist approach will not get a perfect fit but will get a good approximation with fewer parameters~\footnote{Modelling with $A,L$ as tensors will give more parameters and better fits but with limited additional understanding.}.
To obtain parameters with similar values, we choose a value for the magnitudes of fluctuations to be $\frac{m_Q^{(1)}}{32\pi^3} = 1\times 10^{-4}$, $\frac{m_Q^{(2)}}{32\pi^3} = 4\times 10^{-5}$, and then perform a least-squares fit to obtain the other parameters (Fig. \ref{fig:dq_corr}). The standard deviation of the parameters was also determined and the model using these the mean parameter $\pm$ standard deviation is also plotted for each case in Fig. \ref{fig:dq_corr}.

After averaging, we get values for the effective tissue "stiffness" and the shape activity as $L = 10.6 \pm 0.5\mu \mbox{m}^{4} \min^{-1}$ and $A = (2.83 \pm 0.48) \times10^{-3}\min^{-1}$ respectively. 
Note that $\rho^* L \gg D_r$, $m_\rho \gg \rho^* D_r$ and $D \gg D_r$. This is due to the fact that movements of cells relative to one another are small in pupal wing tissue with $D_r = 0.19\mu \mbox{m}^{2} \min^{-1}$ as previously measured in \cite{turley_deep_2023}.

\begin{figure}[!ht]
\centering

	\includegraphics[scale=1]{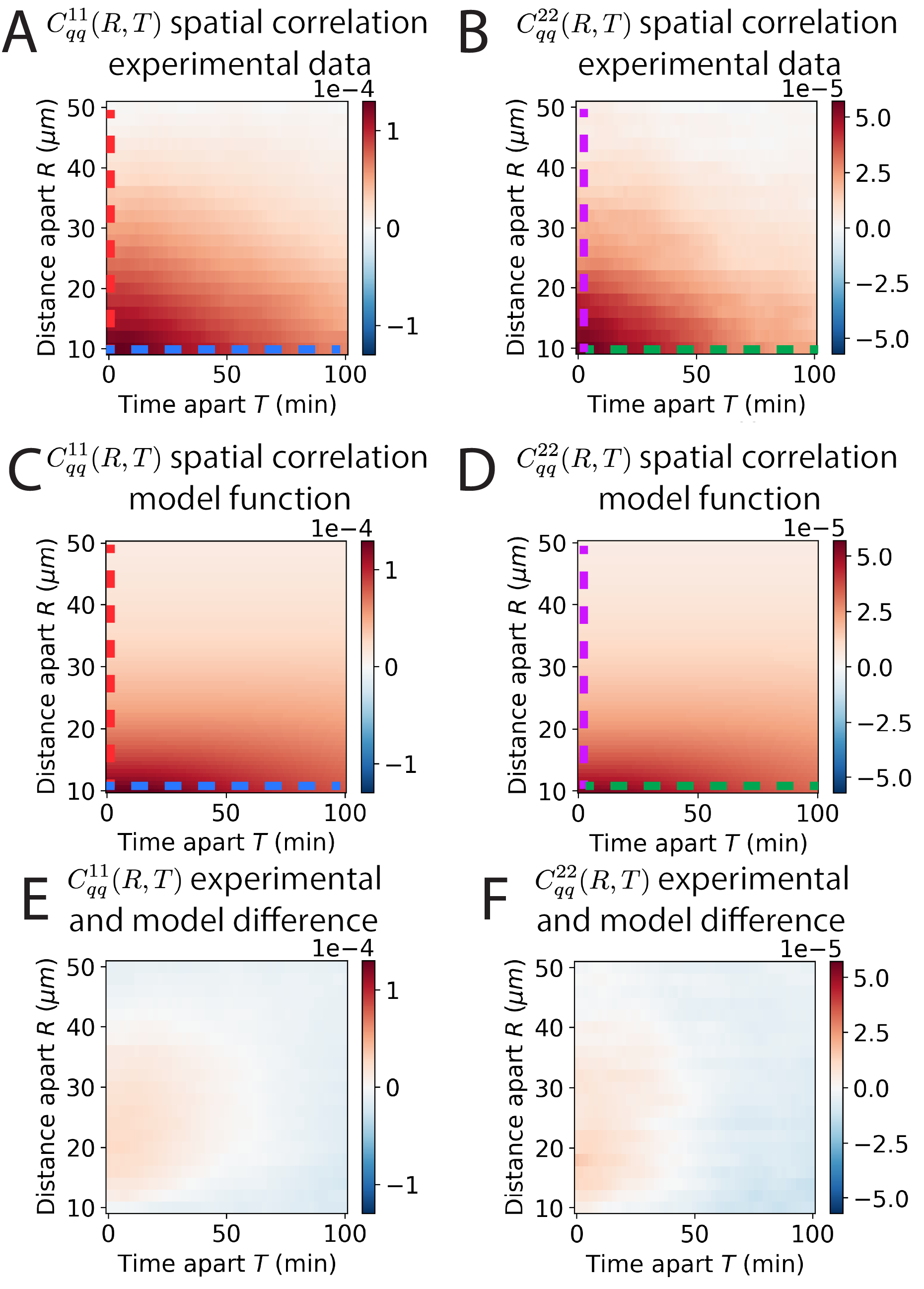} 
	\caption{\justifying \small \sl \textbf{Comparison of averaged model parameters shows close agreement with spatiotemporal correlations in experimental cell shape data.}  Heatmaps of the $\delta Q^{(1)}$ and $\delta Q^{(2)}$ correlations in space and time of A, B) experimental data, C, D) model with fitted parameters and  E, F) the difference between them. Dashed lines on experimental data heatmaps indicate the data used to fit the model in Fig. \ref{fig:dq_corr}.} \label{fig:corr_ST}
\end{figure}

We have plotted the space-time correlations of $\delta Q^{(1)}$ and $\delta Q^{(2)}$ for the experimental data, the model, and their difference in Fig. \ref{fig:corr_ST}. There are small quantitative differences (see Fig. \ref{fig:corr_ST}): the decay of the model is slower than the experimental data and correlations are higher for small $R$. Nevertheless, despite only having a few parameters, the model agrees reasonably well with the data and is good enough for us to extract its effective parameters from our data (Table \ref{tab:parameters}).

\begin{table}
\centering
\begin{tabular}{|p{0.6cm}||p{4.8cm}|p{2.7cm}|}
 \hline
 \multicolumn{2}{|c|}{Parameters of the model} & Value \\
 \hline
 $m_s$ & Magnitude of quenched (structural) density deviation & $4.3\times10^{-4}$ $\mu \mbox{m}^{-4}$\\
 \hline
 $\lambda$ & Quenched (structural) density decay length & 23$\mu \mbox{m}$\\
 \hline
 $D$ & Diffusion constant of annealed density deviations & 7.4 $\mu \mbox{m}^{2} \min^{-1}$\\
 \hline
 $\rho^*$ & Average density & 0.0713 $\mu \mbox{m}^{-2}$\\
 \hline
 $\bar{Q}^{(1)}_0$ & Initial average $Q^{(1)}$ & $7.88\times10^{-3}$\\
 \hline
 $m_{\rho}$ & Annealed density noise magnitude & 109 $\mu \mbox{m}^{-4}$\\
 \hline
 $m_Q^{(1)}$ & $Q^{(1)}$ (PD or AP axes) noise magnitude & $9.92\times10^{-2}$ \\
 \hline
 $m_Q^{(2)}$ & $Q^{(2)}$ ($x=y$ and $x=-y$ axes) noise magnitude & $3.97\times10^{-2}$ \\
 \hline
 $A$ & Active shape change & $(2.83 \pm 0.48) \times10^{-3} \min^{-1}$\\
 \hline
 $b$ & $q$-tensor bulk & 0.0179$\mu \mbox{m}^{2} \min^{-1}$\\
 \hline
 $L$ & effective $q$-tensor "stiffness" & $10.6 \pm 0.5\mu \mbox{m}^{4} \min^{-1}$\\
 \hline
\end{tabular}
\caption{\justifying \small \sl \textbf{Parameters of the model as quantified from experimental data}}
\label{tab:parameters}
\end{table}

\subsubsection{Quantifying changes in tissue dynamics following wound closure}

Next, we test whether the model can distinguish altered tissue states such as those arising in cases of wound healing. Using an ablation laser we wounded the pupal wing epithelium, making circular wounds with diameters of $40 \pm 10 \mu m$ \cite{turley_deep_2024}. As the wound healed, the flies were imaged on the confocal microscope as described above. We continue to  image the tissue for 3 hours until long after the wound has closed to examine if the tissue properties have altered, post wounding in the epithelia. We study wounded epithelial tissue after the {\em wounds have already closed}, so that the same analysis can be applied without needing to account for the presence of the wound. Following wound closure,tissue imaging was continued at the wound site to determine whether the tissue state remained altered in the post-wound tissue. To probe the model, the experiments were performed under different genetic conditions.

We compare (1) wild-type unwounded tissue, (2) wounded wild-type tissue (Control), and (3) unwounded and (4) wounded JNK signalling-inhibited tissue to test whether these perturbations are associated with changes in the inferred model parameters. It is well documented in the literature that expressing a dominant negative form of \textit{bsk}, which prevents JNK signalling, causes cell shape changes to become slower \cite{turley_deep_2024, lee_cdc37_2019, Tafesh-Edwards2020JNKHomeostasis, weavers_injury_2019}. 

\begin{figure}[h!] 
\begin{center}
\includegraphics[scale=1]{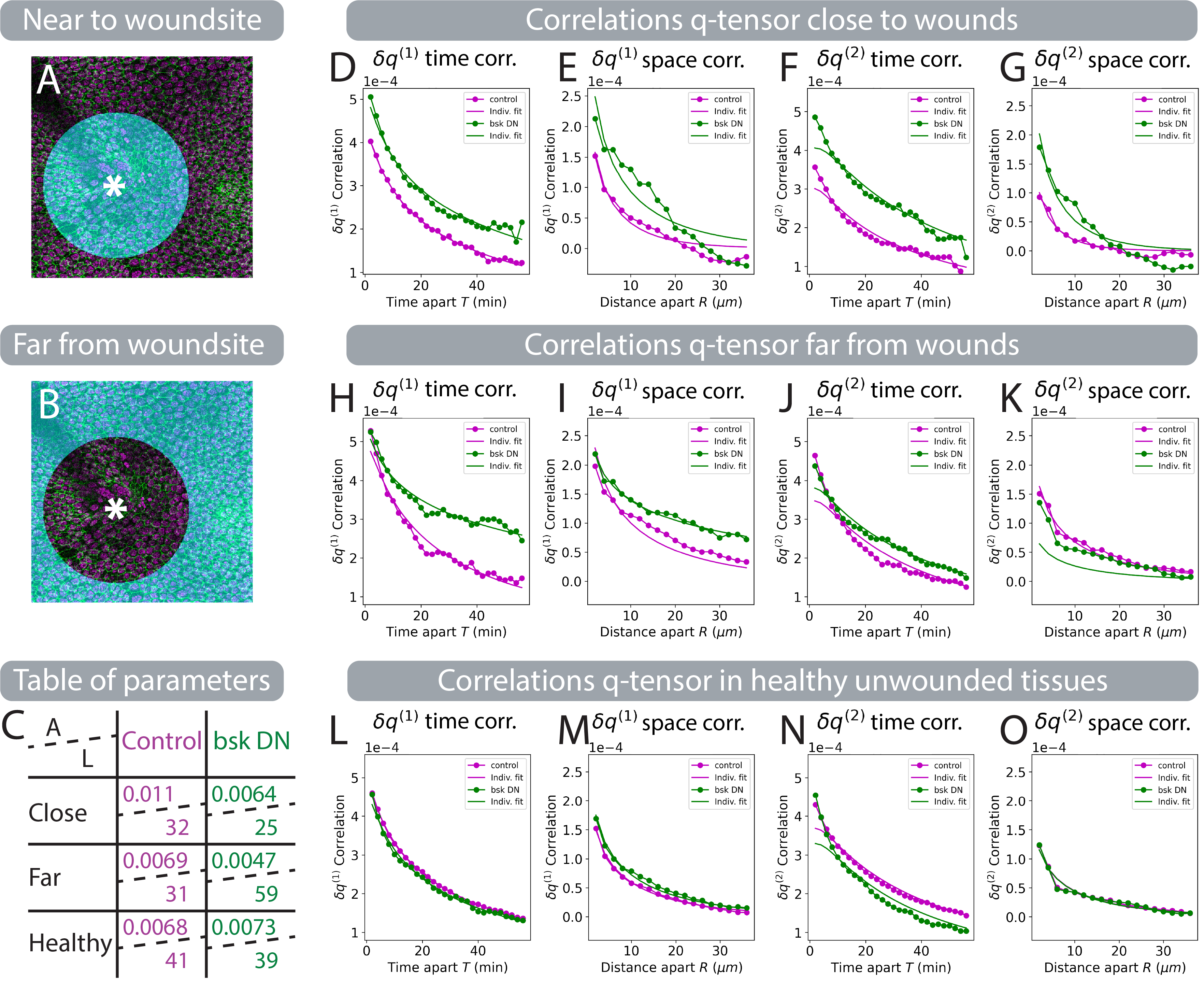} 
\caption{\small \sl \textbf{$q$-tensor fluctuation correlations in JNK knockdown tissue decay more slowly than in controls following wound closure.} A, B) Diagram of cells defined as close to the sealed wound (within 30$\mu m$ from a closed wound) and far from sealed wound (further than 30$\mu m$ from a closed wound). C) Table of mean model parameters from fitting to the different conditions. Experimental correlation functions for control (magenta) and JNK knockdown (green) tissues as a function of time ($R=2$) and space ($T=0$) for $\delta q_1$ and $\delta q_2$. Solid lines show model fits using the mean fitted parameters; shaded regions indicate the standard deviation arising from variation in the fitted model parameters. D-G) Correlations of cells close to a closed wound. H-K) Correlations of cells far from a closed wound. L-O) Correlations of cell in a healthy tissue. }\label{fig:mutant}
\end{center}
\end{figure}

We measure cells in the vicinity of wounds, splitting the tissue into two regions: near the closed wound site (within $30\mu$m) and far from the wound site (Fig.~\ref{fig:mutant}A, B). This binning of $30\mu$m was chosen as this was the region of the tissue where cell shape changes caused by wounding were detectable \cite{turley_deep_2024}. Wounds in control and JNK-inhibited tissues are closed after 90 minutes, so we measure correlations between cells after this time and for the following hour. Due to the smaller datasets, the density correlation functions are too sparse to extract information from, so we focus on $q$-tensor correlations. 

For each condition, we evaluate the correlation functions of the $q$-tensor in space and time, and extract the parameters by comparing experimental data to model using the method described above. As the amplitudes of the $q$-tensor correlations are similar across conditions, we fixed the noise amplitudes $m^{Q^1}$ and $m^{Q^2}$ to the values used for the control tissue and extracted only $A$ and $L$. We set $\frac{m_Q^{(1)}}{32\pi^3} = 1\times 10^{-4}$ and $\frac{m_Q^{(2)}}{32\pi^3} = 4\times 10^{-5}$. The parameters $A$ and $L$ are calculated for each condition and are displayed in the table shown in Fig.~\ref{fig:mutant}C.  We first compare wounded wild-type tissues with unwounded wild-type tissues, we find small changes in $L$ (effective $q$-tensor “stiffness”), with a 22\% reduction in $L$ in tissue close to a closed wound site. This is consistent with a lower effective $q$-tensor "stiffness", which could contribute to wound closure by making the tissue easier to deform. This same reduction in $L$ is also found far from the wound site Fig.~\ref{fig:mutant}H-K.

More significant changes are observed for wounded wild-type tissue (Control) in $A$, the parameter that controls active shape changes. Close to a wound, $A$ is 1.6 times higher than in unwounded tissue. Within the model, this corresponds to a faster relaxation of cell shapes toward the preferred $q$-tensor state. This is reflected in the rapidly decaying correlation functions in Fig.~\ref{fig:mutant}E, G. Cell shape changes are known to contribute to wound closure, and increased active shape remodelling may facilitate faster healing. Interestingly, in regions far from the wound site, where $A$ is similar to healthly tissue. This may indicate that active shape remodelling is spatially restricted to regions immediately adjacent to the wound.

Lastly, we examine differences between wild-type and \textit{bsk}\textsuperscript{DN} tissues. In unwounded tissues of both types, there are minimal differences in $q$-tensor correlation decay (Fig.~\ref{fig:mutant}L–O). However, this is not the case when a wound is introduced. The decay rate of correlations is slower in \textit{bsk}\textsuperscript{DN} tissues, particularly close to the original wound site (Fig.~\ref{fig:mutant}D–K). Here, the tissue appears softer, with a lower $L$.   Most notably however for \textit{bsk}\textsuperscript{DN} mutants, after wounding, near the wound site, $A$ fails to increase to levels comparable to those observed in wild type tissue and remains similar to healthy unwounded tissue. This reduction in active cell shape changes may contribute to the slower wound healing observed in \textit{bsk}\textsuperscript{DN} tissues \cite{turley_deep_2024, lee_cdc37_2019, Tafesh-Edwards2020JNKHomeostasis, weavers_injury_2019}.

In summary, the model distinguishes between passive contributions from effective $q$-tensor stiffness $L$ and active shape remodelling $A$, enabling quantitative comparison across a variety of genetic and wound-healing conditions.

\onecolumngrid

\protect 

\newpage

\twocolumngrid
\medskip

\section{Discussion}

Here we have developed a theory that describes the dynamics of epithelial cells in a tissue, capturing both their mean behaviour and the fluctuations away from this average. Our model consists of a set of stochastic differential equations (SDEs) which can be used to generate correlation functions. These SDEs have a number of parameters that can be related to the mechanical properties of a tissue using the correlation functions. These parameters we have extracted from the experimental data can be used to understand the macroscopic properties of the tissue and to give insights into the interactions between cells at the microscopic scale. This was made possible by our ability to accurately measure, and characterise the shapes and sizes of a large number of cells, using automated deep learning algorithms \cite{turley_deep_2024-1}.

The model may be further refined through the incorporation of additional parameters, thereby achieving a more precise alignment with the experimental data. An improvement could come from removing the assumption that tissue effective "stiffness" is homogeneous. $L_{\gamma\epsilon}$ could be rewritten as $L_{\gamma\epsilon} = L_0 \delta_{\gamma\epsilon} + L_1 \delta_{1 \gamma}\delta_{1 \epsilon}$. Due to the elongation of cells along one axis, it is likely that the tissue could be "stiffer" along this orientation. Adding this parameter could allow for a closer match with the data while maintaining the physical meaning for each parameter. 

For the density correlation we found there was a annealed and quenched component. It's possible that this is also true for cell shape though the timescale maybe longer than our current experiment setup. 

This theoretical model can be used to measure the relative changes mechanical properties of a tissue by observing both the fluctuations and the dynamic response within the tissue via spatio-temporal correlation functions. By measuring tissue properties after the closure of a wound,  we found that tissue effective "stiffness" was reduced in wounded tissues and cell active shape changes increased close to a sealed wound site. After genetically perturbing JNK signalling in the fly wing, we observed spatio-temporal correlations to be decades slower than wild-type tissue and are able to show that this corresponds to a lack of active cell shape changes. The model has been able to distinguish changes in these two parameters which correspond to active and passive properties of the tissue. The ability to distinguish these features may help helpful in determining the biochemical mechanisms driving these effects.

Observing fluctuations is non-invasive compared to laser-recoil experiments and can be done over long length and time-scales. With enough data, the physical properties of a developing tissue could be measured over different time points or locations. Comparison of parameters may show how the properties of tissue change during cancer progression, morphogenesis or wound healing.\\

JT acknowledges a MRC GW4 studentship and Eric and Wendy Schmidt AI in Science Postdoctoral Fellowship. TBL acknowledges the support of Leverhulme Trust Research Project Grant RPG-2016-147. TBL was supported by the UKRI-funded Synthetic Biology Research Centre BrisSynBio (BB/L01386X/1). TBL, JT, IC would like to thank the Isaac Newton Institute for Mathematical Sciences, Cambridge, for support and hospitality during the programme {\em New statistical physics in living matter}, where part of this work was done. This work was supported by EPSRC grants EP/V520287/1, EP/R014604/1 and EP/T031077/1. 

\onecolumngrid

\appendix

\section{Data and code availability}

All data generated or analysed during this study has been deposited in Zenodo at this link https://zenodo.org/records/13819609.

Code used to analyse this data can be found in https://github.com/turleyjm/woundHealing. Main results are in scripts 

\section{Generic microscopic description}

Here we start with microscopic equations of motion for the positions of the cell centres and cell shape ($q$-tensor). We then use them to generate equations for fluctuating fields measuring local cell density and cell shape  by averaging and coarse-graining. We begin by writing the microscopic equations in the most generic form, and then outline the process of coarse-graining in general. In the main text, we have defined a specific minimal microscopic model based on a number of simplifying assumptions and from it obtain the linearised generalised fluctuating hydrodynamic equations.

The generic equations for positions, $\bfr_i(t)$ and shape, $\bmq_i(t)$ of $N$ cells are a combination of deterministic and stochastic terms:
\begin{align} 
 \partial_t \bfr_{i} &= {\bfF}_{i} (\bfr_i,\bmq_i;\{\bmq_j\},\{ \bfr_j\} ) + \bm{\xi}^r_{i} \label{eq:dotr_gen} \\ \partial_t \bmq_{i} &=  \bmG_{i} (\bfr_i,\bmq_i;\{\bmq_j\},\{ \bfr_j\} ) + \bm{\xi}^q_{i} \label{eq:dotq_gen}
\end{align}
where $i \in [1,\ldots,N]$. The noise, $\bm{\xi}^\alpha_i$ have zero mean and  fluctuations $\langle \bm{\xi}_i^{q}(t) \bm{\xi}_j^{q}(t') \rangle = 2 \bmD^q \delta_{ij} \delta(t' - t)$ and $\langle \bm{\xi}_i^{r}(t) \bm{\xi}_j^{r}(t') \rangle = 2 D_r  \bm{\delta}  \delta_{ij} \delta(t' - t)$. 

A class of deterministic velocities we will be particularly interested in our dense tissues are those with "self-interactions" and  "pairwise-interactions":
\beqa
\bfF_{i} &=& \bff_0(\bfr_i,\bmq_i) 
+ \sum_{j\ne i} \bfr_{ij} f_1 (r_{ij}, \bmq_i \cdot \bmq_j, \bfr_{ij}\cdot \bmq_j, \bmq_i \cdot \bmq_j \cdot \bfr_{ij}) + \cdots \nonumber \\
\bmG_i &=& \bm{g}_0 (\bfr_i,\bmq_i) 
+ \sum_{j\ne i} \bmq_j g_1 ( r_{ij}, \bmq_i \cdot \bmq_j, \bfr_{ij}\bfr_{ij} \cdot \bmq_j) + \cdots  
\eeqa
where $\bfr_{ij}=\bfr_i - \bfr_j$ and $r_{ij}= |\bfr_{ij}|$.

We will describe the large scale behaviour by the coarse-grained fields for cell density: $\hat{\rho}(\bfr,t)$, and cell shape: $\hat{\bmQ} (\bfr,t)$ defined as
\beqa
\hat\rho(\bfr, t) &=& \sum_i \delta (\bfr_i(t)- \bfr) =\sum_i \hat \rho_i (\bfr,t) \quad ,  \\
\hat\rho(\bfr, t) \hat{\bmQ} (\bfr, t) &=& \sum_i \bmq_i (t)\delta (\bfr_i(t)- \bfr) =\sum_i \bmq_i \hat \rho_i \quad ,
\eeqa
where $\hat \rho_i (\bfr,t)=  \delta (\bfr_i (t) - \bfr)$.
In a previous study, we also calculated the dynamics of cell shape polarisation~\cite{Olenik2023FluctuationsTissue}. In our pupal wing system, we found that this was negligible and on average zero hence we have not included it here.

Applying the It\^o-calculus~\cite{vankampen, gardiner1985handbook}, we can obtain  stochastic PDEs for the coarse-grained fields,
\beq
\partial_t \hat{\rho} = D_r \bnabla^2 \hat{\rho} + \sum_i \frac{\partial \hat{\rho}}{\partial \bfr_i}
\cdot \bfF_i 
+  \xi^\rho(\bfr,t) 
\quad, \label{Ito:rho} \eeq 
where 
\bem 
\bnabla^2 = {\partial^2 \over \partial \bfr^2} \quad , \quad 
\eem
\bem
\xi^\rho(\bfr,t) =\sum_i {\partial \hat \rho \over \partial \bfr_i}  \cdot \bm{\xi}^r_i \; , \quad
\eem 
and
\beqa \partial_t \left(\hat \rho \hat \bmQ\right)  &=&  D_r \bnabla^2  \left(\hat \rho \hat \bmQ\right)  + \sum_i {\partial  \over \partial \bfr_i} \left(\hat \rho \hat \bmQ\right)
\cdot \left . \bfF_i  \right.  
+ \sum_i {\partial  \over \partial \bmq_i} \left(\hat \rho \hat \bmQ\right) \cdot \left .  \bmG_i \right . + \bm{\xi}^Q (\bfr,t)
\quad ,
\label{Ito:Q}\eeqa
where
\begin{equation}
\bm{\xi}^Q (\mathbf{r},t) = \sum_i \frac{\partial}{\partial \mathbf{r}_i} \left(\hat \rho \, \hat{\mathbf{Q}}\right)
\cdot \bm{\xi}^r_i 
+ \sum_i \frac{\partial}{\partial {\bmq}_i} \left(\hat \rho \, \hat{\mathbf{Q}} \right) 
\bm{\xi}^q_i \, .
\end{equation}

For the deterministic velocities we consider, on simplification, one in general obtains non-local stochastic PDEs for the fields.
These equations become more transparent once one evaluates them for specific functions $\bfF_i$ and $\bmG_i$ as we do in the main text.

\section{Experimental Methods}

The data analysed in the study was used from a previous study \cite{turley_deep_2024} full details can be found there but a summary is given here.

\textit{Drosophila} stocks were raised and maintained on Iberian food according to standard protocols \cite{greenspan_fly_1997}. All crosses were performed at 25°C. The following Drosophila stocks were used: \textit{E-cadherin-GFP} (B\#60584), \textit{Histone2Av-RFP} (B\#23651), \textit{Act-Gal4} (B\#4414), \textit{Tub-Gal$80^{ts}$} (B\#7108) and \textit{UAS-bs$k^{DN}$} (B\#9311). \textit{Drosophila} mutants and transgenic lines were obtained from the Bloomington Stock Centre.

Drosophila pupae were aged to 18h APF at 25°C unless stated otherwise. Dissection, imaging and wounding were all performed as previously described \cite{Weavers2018Long-termPupae}. The time-lapse movies were generated using a SP8 Leica confocal microscope. Each z-stack slice consisted of a 123.26×123.26 $\mu$m image (512×512 pixels) with a slice taken every 0.75 $\mu$m. The flies were wounded on a wide-field microscope, which has a nitrogen-pumped micropoint ablation laser (tuned to 435nm, Andor) attached, before being quickly transferred to the confocal microscope \cite{turley_deep_2024}.

For fly lines where the JNK signalling were to be blocked, we shifted flies to 29°C once they had developed to pre-pupae for 14.75h APF, such that they had developed to the same stage as pupae aged at 18h APF at 25°C. 

\section{Image analysis}

Full details of the image analysis workflow are given in a previous study \cite{turley_deep_2024-1} but a summary is given here.

The individual $3D$ stacks are transformed into a 2D images using a stack focuser \cite{turley_deep_2024-1}. The plugin works by selecting the most in focus pixel for each $(x,y)$. The most in focus pixel is the one with the highest variance of intensity in the surrounding pixels. We apply UNetBoundary to detect the cell boundaries before a watershed algorithm is applied \cite{turley_deep_2024-1}. We can fit polygons to these cells boundaries to calculate their properties such as there position, area and $q$-tensor\cite{Olenik2023FluctuationsTissue}. \\

\section{Correlation functions from experiment}

For the density correlation function we first bin the tissue into $10 \times 10\mu m$ square grids. Then we count the cells with their centroids in a grid and divide by the sum of the cell areas to get the density in the grid. We can take a mean density at each timepoint and use this to find the deviations in the density. We then apply the definition of the density correlation function.

\begin{align}
    C_{\rho\rho}(R, T) &= \langle \delta \rho(\bfr, t) \rho(\bfr', t') \rangle
\end{align}
where the average $\langle ... \rangle$ is over all pairs of grids $R$ apart in distance and $T$ in time.

We use a similar method for the $q$-tensor correlation functions. Here there is no need to first bin the $q$-tensor field, as the $q$-tensor of each individual cell can be used directly. The correlation function is given by
\begin{align}
    C^{\alpha \alpha}_{qq}(R, T) = \frac{1}{N} \sum_{ij}
    \delta q^{(\alpha)}_i(\bfr_i, t_i)
    \delta q^{(\alpha)}_j(\bfr_j, t_j),
\end{align}
where $R = |\bfr_i-\bfr_j|$ and $T = |t_i-t_j|$ are the spatial and temporal separations between pairs of cells. Since using individual cells results in a very large number of possible cell pairs, we randomly sample 5\% of all possible pairs. The sampled pairs are then grouped into bins corresponding to values of $R$ and $T$, and the correlation is averaged within each bin, where $N$ is the number of sampled cell pairs contributing to a given $(R,T)$ bin.

\section{Fitting the model to  correlation data}

Changing the magnitude of fluctuations can give very different values for $A$ and $L$ with equally good fits to each of the 1D data sweeps of the 4 scenarios. By varying the values of $m^{(1)}_{Q}$ and $m^{(2)}_{Q}$ we can get consistent fits for $A$ and $L$ for all the 4 cases. Averaging $A$ and $L$ after finding an optimum values for $m^{(1)}_Q$ and $m_Q^{(2)}$ we get a good approximation of the model parameters. The table shows the parameter values for the least squares fit of the model to the experimental data.
\begin{table}[h!]
\centering
\begin{tabular}{|p{1.7cm}||p{1.4cm}|p{1.4cm}|p{1.4cm}|p{1.4cm}| }
 \hline
 Parameter & $C^{11}_{qq}(10, t)$ & $C^{11}_{qq}(\bfr, 0)$ & $C^{22}_{qq}(10, t)$ & $C^{22}_{qq}(\bfr, 0)$ \\
 \hline
 $A$ & 0.00326 & 0.00236 & 0.00337 & 0.00234 \\
 \hline
 $L$ & 10.4 & 11.4 & 10.5 & 10.1 \\
 \hline
\end{tabular}
\end{table}

\section{Some Calculational details}

\subsection{Coarse-graining the equations}
The second term of the equation for the $Q$-tensor, eqn. (\ref{eq:dotQExpand}) will require computing 
\begin{align} 
-\sum_i \left . \frac{\partial U}{\partial q_{i\alpha\beta}} \delta(\bfr - \bfr_i(t)) \right . =  -\sum_i \frac{\partial U}{\partial q_{i\alpha\beta}} \hat \rho_i \label{eq:second term_q}
\end{align}

$\frac{\partial U}{\partial q_{i\alpha\beta}}$ can be computed as 
\begin{align} 
\frac{\partial U}{\partial q_{i\alpha\beta}} &= \sum_j W(\bfr_i - \bfr_j) \partial_{q_{i\alpha\beta}} \Tr(\bmq_i \bmq_j) \nonumber \\
&= \sum_j 2q_{j\alpha\beta} W_2(\bfr_i - \bfr_j) \label{eq:dU/dq}
\end{align}
since $\partial_{\bmq_i} \Tr(\bmq_i \bmq_j) = 2\bmq_j$.  

Substituting (\ref{eq:dU/dq}) in (\ref{eq:second term_q}) we get 
\begin{align} 
-\sum_i \frac{\partial U}{\partial q_{i\alpha\beta}} \hat \rho_i &= -2\sum_{ij}  q_{j\alpha\beta} W_2(\bfr_i - \bfr_j)\delta(\bfr - \bfr_i(t)) \nonumber \\
&=   -2 \int_{\mathbb{R}^2}  d^2 r' \, \, \hat \rho (\bfr',t)\hat Q (\bfr',t) W_2(\bfr-\bfr')\hat \rho (\bfr, t) \\
&= -2\int_{\mathbb{R}^2} d^2 r' \, [\bar{\rho}^2\bar{Q}_{\alpha\beta} + \bar{\rho}^2\delta Q_{\alpha\beta}(\bfr')  
+ \bar{\rho}\bar{Q}_{\alpha\beta}\delta\rho(\bfr) + \bar{\rho}\bar{Q}_{\alpha\beta}\delta\rho(\bfr')] W_2(\bfr - \bfr')  \nonumber 
\end{align}
upon local averaging and linearization.

By performing Taylor expansions, $\delta \rho (\bfr')= \delta \rho (\bfr) + (\bfr'-\bfr) \cdot \nabla\delta \rho (\bfr') + ...$, and $\delta  Q_{\alpha\beta} (\bfr')= \delta Q_{\alpha\beta} (\bfr) + (\bfr'-\bfr) \cdot \nabla\delta  Q_{\alpha\beta} (\bfr') + ...$
we get
\begin{align} 
-\sum_i \left . \frac{\partial U}{\partial q_{i\alpha\beta}} \hat \rho_i \right . &= 2b(\bar{\rho}^2\bar{Q}_{\alpha\beta} + \bar{\rho}^2\delta Q_{\alpha\beta}(\bfr) + 2\bar{\rho}\bar{Q}_{\alpha\beta}\delta\rho(\bfr)) 
+ L_{\gamma\epsilon}\partial_{\gamma} \partial_{\epsilon}(\bar{\rho}^2\delta Q_{\alpha\beta}(\bfr) + \bar{\rho}\bar{Q}_{\alpha\beta}\delta\rho(\bfr)) \; .
\end{align}
We get the parameters from integrals over the energy of shape interactions,
\begin{align} 
b &= - \int_{\mathbb{R}^2} d^2 r \, W(\bfr)  \\
L_{\gamma\epsilon} &= -\int_{\mathbb{R}^2} d^2 r \, \left( r_\gamma r_\epsilon \right)  \, W(\bfr) 
\end{align}
The shape interactions generate a bulk passive isotropic stress (pressure) in the tissue, as well as a tissue stiffness. $b$ is a measure of the pressure and $L_{\gamma\epsilon}$ is a measure of the stiffness of the tissue. 

\subsection{Calculating correlations}

We compute the density-correlation functions 
for $t'>t$ as follows: 
\begin{equation}
  \begin{split}
    & \langle \delta\rho(\bfr, t) \delta\rho(\bfr', t') \rangle = - \int_{\mathbb{R}^2} \frac{d^2 k}{(2\pi)^2} \int_{\mathbb{R}^2} \frac{d^2 k' }{(2\pi)^2} e^{i\bfk \cdot \bfr+i\bfk' \cdot \bfr'} 
    \left \langle \int_{-\infty}^t d\tau \int_{-\infty}^{t'} d\tau' k_\gamma k'_\epsilon \tilde{\xi}_{\gamma}^{\rho}(\bfk,\tau) \tilde{\xi}_{\epsilon}^{\rho}(\bfk',\tau') e^{-Dk^2(t-\tau)-Dk'^2(t'-\tau')} \right \rangle \\
  \end{split}
\end{equation}
From eqn. (\ref{eq:rho-noise}), it follows that $\langle\tilde{\xi}_{\gamma}^{\rho}(\bfk,\tau) \tilde{\xi}_{\epsilon}^{\rho}(\bfk',\tau') \rangle = m_{\rho} \delta_{\gamma\epsilon} \delta(t - t') \delta(\bfk + \bfk')$, hence we obtain 
\begin{align}
  \langle \delta\rho(\bfr, t) \delta\rho(\bfr', t') \rangle &= \frac{m_{\rho}}{2D(2\pi)^4} \int_{\mathbb{R}^2} d^2 k \, e^{-D(t'-t) k^2 - \frac{(\bfr - \bfr')^2}{4D(t'-t)}} \\
  &= \frac{m_{\rho}}{32D^2\pi^3 |t-t'|} \, e^{- \frac{|\bfr-\bfr'|^2}{4DT}} \quad . \nonumber
\end{align}

Similarly, we calculate the shape correlations as follows. 
From the general solution of the equation for $Q^{(1)}(\bfr, t)$ we obtain
\begin{align}
    \langle \delta Q^{(1)}(\bfr, t) &\delta Q^{(1)}(\bfr', t') \rangle  = \frac{1}{(2\pi)^4} \int_{\mathbb{R}^2} d^2 k \int_{\mathbb{R}^2} d^2 k' e^{i\bmk \cdot \bfr+i\bmk' \cdot \bfr'} 
    \int_{-\infty}^t d\tau \int_{-\infty}^{t'} d\tau' \langle \tilde{\xi}_{Q}^{(1)}(\bmk, \tau)\tilde{\xi}_{Q}^{(1)}(\bmk', \tau') \rangle \nonumber\\
    & \times e^{-\rho^*(B + L^*k^2)(t-\tau)-(B + L^*k'^2)(t'-\tau')} \quad .
\end{align}

From eqn. (\ref{eq:Q-alpha-noise}), it follows that  $\langle\tilde{\xi}_{Q}^{(1)}(\bmk,\tau) \tilde{\xi}_{Q}^{(1)}(\bmk',\tau') \rangle = m_Q^{(1)}\delta(t - t') \delta(\bmk + \bmk')$ hence
\begin{align}
  \langle \delta Q^{(1)}(\bfr, t) &\delta Q^{(1)}(\bfr', t') \rangle = \frac{m_Q^{(1)}}{32\pi^4} \int_{\mathbb{R}^2}  \frac{d^2 k}{B + L^*k^2}\, e^{i\bmk \cdot (\bfr - \bfr')}  e^{-(B + L^*k^2)(t'-t)} \quad . & \nonumber
\end{align}

We perform the integral by transforming to polar coordinates, defining $R \equiv |\bfr-\bfr'|$, also taking $T \equiv t' - t$. Finally, we apply Bessel's first integral to obtain
\begin{align}
 C^{11}_{qq}(R, T)   &= \frac{m_{Q}^{(1)}}{32\pi^3\rho^*} \int_0^\infty dk \frac{k J_0(kR)}{B + \rho^*Lk^2} e^{-(B + \rho^*Lk^2)T} \quad . 
\end{align}
An identical analysis can be done for $Q^{(2)}(\bfr, t)$ and $C^{22}_{qq}(R, T)$.

\twocolumngrid

 
\bibliographystyle{apsrev4-1}
\bibliography{References} 

@article{Tetley2019TissueHealing,
    title = {{Tissue fluidity promotes epithelial wound healing}},
    year = {2019},
    journal = {Nature Physics},
    author = {Tetley, Robert J. and Staddon, Michael F. and Heller, Davide and Hoppe, Andreas and Banerjee, Shiladitya and Mao, Yanlan},
    number = {11},
    month = {11},
    pages = {1195--1203},
    volume = {15},
    publisher = {Nature Publishing Group},
    doi = {10.1038/s41567-019-0618-1},
    issn = {17452481}
}

@article{Lenne2022SculptingTransitions,
    title = {{Sculpting tissues by phase transitions}},
    year = {2022},
    journal = {Nature Communications},
    author = {Lenne, Pierre François and Trivedi, Vikas},
    number = {1},
    month = {12},
    volume = {13},
    publisher = {Nature Research},
    doi = {10.1038/s41467-022-28151-9},
    issn = {20411723},
    pmid = {35115507}
}

@article{Mongera2018AElongation,
    title = {{A fluid-to-solid jamming transition underlies vertebrate body axis elongation}},
    year = {2018},
    journal = {Nature},
    author = {Mongera, Alessandro and Rowghanian, Payam and Gustafson, Hannah J. and Shelton, Elijah and Kealhofer, David A. and Carn, Emmet K. and Serwane, Friedhelm and Lucio, Adam A. and Giammona, James and Camp{\`{a}}s, Otger},
    number = {7723},
    month = {9},
    pages = {401--405},
    volume = {561},
    publisher = {Nature Publishing Group},
    doi = {10.1038/s41586-018-0479-2},
    issn = {14764687},
    pmid = {30185907}
}

@article{rustarazo-calvo_adhesion-driven_2026,
	title = {Adhesion-driven rigidity transition decoupled from density-driven jamming triggers epithelial organization in embryonic tissues},
	issn = {1745-2473, 1745-2481},
	url = {https://www.nature.com/articles/s41567-026-03276-6},
	doi = {10.1038/s41567-026-03276-6},
	urldate = {2026-06-09},
	journal = {Nature Physics},
	author = {Rustarazo-Calvo, Laura and Pallares-Cartes, Cristina and Aguirre-Tamaral, Adrián and Floris, Elisa and Hingerl, Maximilian and Autorino, Camilla and Khan, Arif Ul Maula and Corominas-Murtra, Bernat and Petridou, Nicoletta I.},
	month = jun,
	year = {2026},
}

@misc{staddon_cell_2026,
	title = {Cell proliferation maintains cell area polydispersity in the growing fruit fly wing epithelium},
	url = {http://arxiv.org/abs/2601.14509},
	doi = {10.48550/arXiv.2601.14509},
	urldate = {2026-06-09},
	publisher = {arXiv},
	author = {Staddon, Michael F. and Dye, Natalie A. and Popović, Marko and Jülicher, Frank},
	month = jan,
	year = {2026},
	note = {arXiv:2601.14509 [physics.bio-ph]},
}

@article{fuhrmann_active_2024,
	title = {Active shape programming drives {Drosophila} wing disc eversion},
	journal = {Science AdvAnceS},
	author = {Fuhrmann, Jana F and Krishna, Abhijeet and Paijmans, Joris and Duclut, Charlie and Cwikla, Greta and Eaton, Suzanne and Popović, Marko and Jülicher, Frank and Modes, Carl D and Dye, Natalie A},
	year = {2024},
}

@article{Tetley2018TheFluidity,
    title = {{The same but different: Cell intercalation as a driver of tissue deformation and fluidity}},
    year = {2018},
    journal = {Philosophical Transactions of the Royal Society B: Biological Sciences},
    author = {Tetley, Robert J. and Mao, Yanlan},
    number = {1759},
    volume = {373},
    publisher = {Royal Society Publishing},
    doi = {10.1098/rstb.2017.0328},
    issn = {14712970},
    pmid = {30249777}
}

@article{Hirashima2017CellularMorphogenesis,
    title = {{Cellular Potts modeling of complex multicellular behaviors in tissue morphogenesis}},
    year = {2017},
    journal = {Development Growth and Differentiation},
    author = {Hirashima, Tsuyoshi and Rens, Elisabeth G. and Merks, Roeland M.H.},
    number = {5},
    month = {6},
    pages = {329--339},
    volume = {59},
    publisher = {Blackwell Publishing},
    doi = {10.1111/dgd.12358},
    issn = {1440169X},
    pmid = {28593653}
}

@misc{leong_critical_2025,
	title = {Critical phenomenon underlies de novo luminogenesis during mammalian follicle development},
	publisher = {bioRxiv},
	author = {Leong, Kim Whye and Lou, Yuting and Biswas, Arikta and Tan, Jue Yu Kelly and Ng, Boon Heng and Lu, Xixun and Teo, Xin Ping Joan and Lu, Thong Beng and Bevilacqua, Carlo and Bonne, Isabelle and Prevedel, Robert and Hiraiwa, Tetsuya and Chan, Chii Jou},
	year = {2025},
}

@article{turley_iscience_2022,
	title = {{iScience} {What} good is maths in studies of wound healing?},
	volume = {25},
	copyright = {All rights reserved},
	url = {https://doi.org/10.1016/j.isci.},
	journal = {ISCIENCE},
	author = {Turley, Jake and Chenchiah, Isaac V and Liverpool, Tanniemola B and Weavers, Helen and Martin, Paul},
	year = {2022},
	pages = {104778},
}

@misc{tozluolu_planar_2019,
	title = {Planar differential growth rates determine the position of folds in complex epithelia},
	url = {http://biorxiv.org/lookup/doi/10.1101/515528},
	doi = {10.1101/515528},
	urldate = {2024-10-25},
	author = {Tozluolu, Melda and Duda, Maria and Kirkland, Natalie J. and Barrientos, Ricardo and Burden, Jemima J. and Muñoz, José J. and Mao, Yanlan},
	month = jan,
	year = {2019},
}

@article{manning_rigidity_2024,
	title = {Rigidity in mechanical biological networks},
	volume = {34},
	issn = {09609822},
	url = {https://linkinghub.elsevier.com/retrieve/pii/S0960982224009175},
	doi = {10.1016/j.cub.2024.07.014},
	number = {20},
	urldate = {2024-10-24},
	journal = {Current Biology},
	author = {Manning, M. Lisa},
	month = oct,
	year = {2024},
	pages = {R1024--R1030},
}

@article{park_unjamming_2015,
	title = {Unjamming and cell shape in the asthmatic airway epithelium},
	volume = {14},
	issn = {1476-1122, 1476-4660},
	url = {https://www.nature.com/articles/nmat4357},
	doi = {10.1038/nmat4357},
	number = {10},
	urldate = {2024-10-25},
	journal = {Nature Materials},
	author = {Park, Jin-Ah and Kim, Jae Hun and Bi, Dapeng and Mitchel, Jennifer A. and Qazvini, Nader Taheri and Tantisira, Kelan and Park, Chan Young and McGill, Maureen and Kim, Sae-Hoon and Gweon, Bomi and Notbohm, Jacob and Steward Jr, Robert and Burger, Stephanie and Randell, Scott H. and Kho, Alvin T. and Tambe, Dhananjay T. and Hardin, Corey and Shore, Stephanie A. and Israel, Elliot and Weitz, David A. and Tschumperlin, Daniel J. and Henske, Elizabeth P. and Weiss, Scott T. and Manning, M. Lisa and Butler, James P. and Drazen, Jeffrey M. and Fredberg, Jeffrey J.},
	month = oct,
	year = {2015},
	pages = {1040--1048},
}

@article{gomez_highly_2024,
  title={Highly dynamic mechanical transitions in embryonic cell populations during Drosophila gastrulation},
  author={Gomez, Juan Manuel and Bevilacqua, Carlo and Thayambath, Abhisha and Heriche, Jean-Karim and Leptin, Maria and Belmonte, Julio M and Prevedel, Robert},
  journal={Nature Communications},
  volume={16},
  number={1},
  pages={6473},
  year={2025},
  publisher={Nature Publishing Group UK London}
}

@article{Etournay2015InterplayWing,
    title = {{Interplay of cell dynamics and epithelial tension during morphogenesis of the Drosophila pupal wing}},
    year = {2015},
    journal = {ELife},
    author = {Etournay, Raphä and Popov{\'{i}}, Marko and Merkel, Matthias and Nandi, Amitabha and Blasse, Corinna and Aigouy, BenoˆıtBenoˆıt and Brandl, Holger and Myers, Gene and Salbreux, Guillaume and Eaton, Suzanne},
    doi = {10.7554/eLife.07090.001}
}

@article{ishihara_cells_2017,
	title = {From cells to tissue: {A} continuum model of epithelial mechanics},
	volume = {96},
	copyright = {https://link.aps.org/licenses/aps-default-license},
	issn = {2470-0045, 2470-0053},
	shorttitle = {From cells to tissue},
	url = {https://link.aps.org/doi/10.1103/PhysRevE.96.022418},
	doi = {10.1103/PhysRevE.96.022418},
	number = {2},
	urldate = {2024-10-25},
	journal = {Physical Review E},
	author = {Ishihara, Shuji and Marcq, Philippe and Sugimura, Kaoru},
	month = aug,
	year = {2017},
	pages = {022418},
}

@article{murisic_discrete_2015,
	title = {From {Discrete} to {Continuum} {Models} of {Three}-{Dimensional} {Deformations} in {Epithelial} {Sheets}},
	volume = {109},
	issn = {00063495},
	url = {https://linkinghub.elsevier.com/retrieve/pii/S0006349515005044},
	doi = {10.1016/j.bpj.2015.05.019},
	number = {1},
	urldate = {2024-10-25},
	journal = {Biophysical Journal},
	author = {Murisic, Nebojsa and Hakim, Vincent and Kevrekidis, Ioannis G. and Shvartsman, Stanislav Y. and Audoly, Basile},
	month = jul,
	year = {2015},
	pages = {154--163},
}

@article{andralojc_dynamics_2026,
	title = {Dynamics of {Wound} {Closure} in {Living} {Nematic} {Epithelia}},
	volume = {136},
	issn = {0031-9007, 1079-7114},
	url = {https://link.aps.org/doi/10.1103/8871-8m6c},
	doi = {10.1103/8871-8m6c},
	number = {15},
	urldate = {2026-06-09},
	journal = {Physical Review Letters},
	author = {Andralojc, Henry and Turley, Jake and Weavers, Helen and Martin, Paul and Chenchiah, Isaac V. and Bennett, Rachel R. and Liverpool, Tanniemola B.},
	month = apr,
	year = {2026},
	pages = {158402},
}

@misc{ioratim-uba_exotic_2025,
	title = {Exotic rheology of materials with active rearrangements},
	url = {http://arxiv.org/abs/2508.19844},
	doi = {10.48550/arXiv.2508.19844},
	urldate = {2026-06-09},
	publisher = {arXiv},
	author = {Ioratim-Uba, Aondoyima and Liverpool, Tanniemola B. and Henkes, Silke},
	month = aug,
	year = {2025},
	note = {arXiv:2508.19844 [cond-mat.soft]},
}

@article{turley_deep_2024,
	title = {Deep learning reveals a damage signalling hierarchy that coordinates different cell behaviours driving wound re-epithelialisation},
	volume = {151},
	copyright = {http://creativecommons.org/licenses/by/4.0},
	issn = {0950-1991, 1477-9129},
	url = {https://journals.biologists.com/dev/article/151/18/dev202943/362123/Deep-learning-reveals-a-damage-signalling},
	doi = {10.1242/dev.202943},
	number = {18},
	urldate = {2024-10-25},
	journal = {Development},
	author = {Turley, Jake and Robertson, Francesca and Chenchiah, Isaac V. and Liverpool, Tanniemola B. and Weavers, Helen and Martin, Paul},
	month = sep,
	year = {2024},
	pages = {dev202943},
}

@article{turley_deep_2024-1,
	title = {Deep learning for rapid analysis of cell divisions in vivo during epithelial morphogenesis and repair},
	volume = {12},
	copyright = {All rights reserved},
	issn = {2050-084X},
	url = {https://elifesciences.org/articles/87949},
	doi = {10.7554/eLife.87949},
	urldate = {2024-10-25},
	journal = {eLife},
	author = {Turley, Jake and Chenchiah, Isaac V and Martin, Paul and Liverpool, Tanniemola B and Weavers, Helen},
	month = sep,
	year = {2024},
	pages = {RP87949},
}

@article{weavers_injury_2019,
	title = {Injury {Activates} a {Dynamic} {Cytoprotective} {Network} to {Confer} {Stress} {Resilience} and {Drive} {Repair}},
	volume = {29},
	issn = {09609822},
	doi = {10.1016/j.cub.2019.09.035},
	number = {22},
	journal = {Current Biology},
	publisher = {Cell Press},
	author = {Weavers, Helen and Wood, Will and Martin, Paul},
	month = nov,
	year = {2019},
	pages = {3851--3862.e4},
}

@article{paci_forced_2021,
	title = {Forced into shape: {Mechanical} forces in {Drosophila} development and homeostasis},
	volume = {120},
	issn = {10963634},
	doi = {10.1016/j.semcdb.2021.05.026},
	journal = {Seminars in Cell and Developmental Biology},
	publisher = {Elsevier Ltd},
	author = {Paci, Giulia and Mao, Yanlan},
	month = dec,
	year = {2021},
	pages = {160--170},
}

@article{athilingam_mechanics_2021,
	title = {Mechanics of epidermal morphogenesis in the {Drosophila} pupa},
	volume = {120},
	issn = {10963634},
	url = {https://doi.org/10.1016/j.semcdb.2021.06.008},
	journal = {Seminars in Cell and Developmental Biology},
	publisher = {Elsevier Ltd},
	author = {Athilingam, Thamarailingam and Tiwari, Prabhat and Toyama, Yusuke and Saunders, Timothy E.},
	month = dec,
	year = {2021},
	pages = {171--180},
}

@article{tetley_tissue_2019,
	title = {Tissue fluidity promotes epithelial wound healing},
	volume = {15},
	issn = {17452481},
	url = {https://doi.org/10.1038/s41567-019-0618-1},
	number = {11},
	journal = {Nature Physics},
	publisher = {Nature Publishing Group},
	author = {Tetley, Robert J. and Staddon, Michael F. and Heller, Davide and Hoppe, Andreas and Banerjee, Shiladitya and Mao, Yanlan},
	month = nov,
	year = {2019},
	pages = {1195--1203},
}

@article{lee_cdc37_2019,
	title = {Cdc37 is essential for {JNK} pathway activation and wound closure in {Drosophila}},
	volume = {30},
	issn = {19394586},
	doi = {10.1091/mbc.E18-12-0822},
	number = {21},
	journal = {Molecular Biology of the Cell},
	publisher = {American Society for Cell Biology},
	author = {Lee, Chan Wool and Kwon, Young Chang and Lee, Youngbin and Park, Min Yoon and Choe, Kwang Min},
	month = oct,
	year = {2019},
	pages = {2651--2658},
}

@article{Tafesh-Edwards2020JNKHomeostasis,
  title={JNK signaling in Drosophila immunity and homeostasis},
  author={Tafesh-Edwards, Ghada and Eleftherianos, Ioannis},
  journal={Immunology Letters},
  volume={226},
  pages={7--11},
  year={2020},
  publisher={Elsevier}
}

@article{Weavers2016SystemsGradient,
    title = {{Systems Analysis of the Dynamic Inflammatory Response to Tissue Damage Reveals Spatiotemporal Properties of the Wound Attractant Gradient}},
    year = {2016},
    journal = {Current Biology},
    author = {Weavers, Helen and Liepe, Juliane and Sim, Aaron and Wood, Will and Martin, Paul and Stumpf, Michael P.H.},
    number = {15},
    month = {8},
    pages = {1975--1989},
    volume = {26},
    publisher = {Cell Press},
    doi = {10.1016/j.cub.2016.06.012},
    issn = {09609822},
    pmid = {27426513}
}

@article{Weavers2018Long-termPupae,
    title = {{Long-term in vivo tracking of inflammatory cell dynamics within drosophila pupae}},
    year = {2018},
    journal = {Journal of Visualized Experiments},
    author = {Weavers, Helen and Franz, Anna and Wood, Will and Martin, Paul},
    number = {136},
    month = {6},
    volume = {2018},
    publisher = {Journal of Visualized Experiments},
    doi = {10.3791/57871},
    issn = {1940087X},
    pmid = {29985351}
}

@article{Olenik2023FluctuationsTissue,
    title = {{Fluctuations of cell geometry and their nonequilibrium thermodynamics in living epithelial tissue}},
    year = {2023},
    journal = {Physical Review E},
    author = {Olenik, M. and Turley, J. and Cross, S. and Weavers, H. and Martin, P. and Chenchiah, I. V. and Liverpool, T. B.},
    number = {1},
    month = {1},
    volume = {107},
    publisher = {American Physical Society},
    doi = {10.1103/PhysRevE.107.014403},
    issn = {24700053},
    pmid = {36797912},
    arxivId = {2201.07154}
}

@article{Tinevez2017TrackMate:Tracking,
    title = {{TrackMate: An open and extensible platform for single-particle tracking}},
    year = {2017},
    journal = {Methods},
    author = {Tinevez, Jean Yves and Perry, Nick and Schindelin, Johannes and Hoopes, Genevieve M. and Reynolds, Gregory D. and Laplantine, Emmanuel and Bednarek, Sebastian Y. and Shorte, Spencer L. and Eliceiri, Kevin W.},
    month = {2},
    pages = {80--90},
    volume = {115},
    publisher = {Academic Press Inc.},
    doi = {10.1016/j.ymeth.2016.09.016},
    issn = {10959130},
    pmid = {27713081}
}

@phdthesis{turley_deep_2023,
	title = {Deep learning and mathematical analysis  of wound healing in flies},
	school = {University of Bristol},
	author = {Turley, Jake},
	month = jun,
	year = {2023},
}

@article{dean1996langevin,
  title={Langevin equation for the density of a system of interacting Langevin processes},
  author={Dean, David S},
  journal={Journal of Physics A: Mathematical and General},
  volume={29},
  number={24},
  pages={L613--L617},
  year={1996}
}

@book{vankampen,
address = {Amsterdam},
author = {Kampen, N G Van},
publisher = {Elsevier},
title = {{Stochastic processes in physics and chemistry}},
year = {2001}
}

@article{gardiner1985handbook,
  title={Handbook of stochastic methods for physics, chemistry and the natural sciences},
  author={Gardiner, Crispin W},
  journal={Springer series in synergetics},
  year={1985},
  publisher={Springer Berlin Heidelberg}
}

@book{greenspan_fly_1997,
	title = {Fly {Pushing}: {The} {Theory} and {Practice} of {Drosophila} {Genetics}},
	volume = {2nd},
	publisher = {Cold Spring Harbor Press},
	author = {Greenspan, J, R},
	year = {1997},
}


\end{document}